\documentclass[a4paper,11pt]{article}

\usepackage{float}
\usepackage{graphicx}
\usepackage{caption}
\usepackage{subcaption}
\usepackage{slashed,nicefrac}

\usepackage{amsmath}

\usepackage{mathtools}

\usepackage{fullpage}
\usepackage[T1]{fontenc}

\usepackage{amssymb,amsmath,cite}

\usepackage{xcolor,bbold,comment}

\usepackage{amsthm}

\begin{document}

\newcommand{\be}{\begin{eqnarray}}
\newcommand{\ee}{\end{eqnarray}}

\def\thefootnote{\arabic{footnote}}
\setcounter{footnote}{0}

\def\nn{\nonumber}

\allowdisplaybreaks

\begin{titlepage}
\thispagestyle{empty}

\begin{flushright}
	\hfill{\ } 
\end{flushright}
				
\vspace{35pt}
				
\begin{center} 
	{\Large{\bf 
	Anti-brane uplift instability from goldstino condensation  
	}} 
									
	\vspace{50pt}
							
	{Gianguido~Dall'Agata, Maxim~Emelin, Fotis~Farakos and Matteo~Morittu}
							
	\vspace{25pt}
							
	{
		{\it Dipartimento di Fisica e Astronomia ``Galileo Galilei''\\
			Universit\`a di Padova, Via Marzolo 8, 35131 Padova, Italy}
										
		\vspace{15pt}
										
		{\it  INFN, Sezione di Padova \\
		Via Marzolo 8, 35131 Padova, Italy}
		}
								
\vspace{40pt}
								
{ABSTRACT} 
\end{center}

We investigate the possible appearance of composite states of the goldstino in models with four-dimensional non-linear supersymmetry and we provide a description of their dynamics in terms of a K\"ahler potential and a superpotential. Our analysis shows that the critical point corresponding to the Volkov--Akulov model is unstable. Similarly, we find that the uplifted stable de Sitter critical point of the KKLT model is shifted and acquires a tachyonic instability. Our findings indicate the existence of a potentially dangerous instability shared by all anti-brane uplifts.

\vspace{10pt}
			
\bigskip
			
\end{titlepage}

\numberwithin{equation}{section}

\baselineskip 6.1 mm

\tableofcontents

\newpage

\section{Introduction}

One of the central ingredients in typical string theory de Sitter constructions is the use of anti-D3-branes to uplift an AdS vacuum to a de Sitter critical configuration \cite{Kachru:2003aw,Balasubramanian:2005zx,Conlon:2005ki,Kallosh:2018nrk,Bento:2021nbb,Bena:2022cwb}. Even though the end result of such uplift is often challenged \cite{Danielsson:2018ztv}, and the existence of tachyons or loss of criticality is implied \cite{Obied:2018sgi,Andriot:2018wzk,Garg:2018reu,Ooguri:2018wrx}, there is no conclusive indication that the four-dimensional effective field theory suffers from the alleged instabilities.
(Recent selected possible issues from a ten-dimensional perspective are further reported in the articles \cite{Moritz:2017xto,Sethi:2017phn,Gao:2020xqh,Hamada:2019ack,Carta:2019rhx,Gautason:2019jwq,Hamada:2018qef,Junghans:2022exo}). 
If one assumes that there is a consistent low energy effective field theory that can incorporate the anti-D3-brane uplift, then one expects to be able to embed it within 4D $N=1$ Supergravity with the use of a nilpotent chiral superfield that breaks supersymmetry and provides the uplift \cite{Lindstrom:1979kq,Kapustnikov:1981de,Samuel:1982uh,Bergshoeff:2015tra,Cribiori:2017ngp,DallAgata:2016syy}. 
Indeed, the embedding of the KKLT-type uplift within Supergravity coupled to a nilpotent superfield was discussed in \cite{Ferrara:2014kva,Bergshoeff:2015jxa}, and the underlying non-linear supersymmetry of the anti-brane uplift was made manifest. More generally, the appearance of a goldstino sector on anti-Dp-brane worldvolumes was also investigated in \cite{Bandos:2015xnf,Dasgupta:2016prs,Cribiori:2019bfx,Cribiori:2020bgt}. 

If the aforementioned simple de Sitter constructions are indeed unstable, one of the cleanest ways to see this would be to uncover this instability in the low energy 4D $N=1$ supergravity description. To support this approach, let us observe that there do exist hints that the Volkov--Akulov (VA) model coupled to pure 4D $N=1$ Supergravity \cite{Volkov:1973ix,Deser:1977uq} is inherently unstable. 
This is seen by considering a real scalar representing a gravitino condensate $\langle \psi_m \psi^m \rangle \sim \sigma (x)$, and studying the effective theory \`a la Nambu--Jona-Lasinio \cite{Nambu:1961tp,Nambu:1961fr} (also similarly to composite Higgs models \cite{Bardeen:1989ds}) for the four-gravitini interaction. 
Such an analysis has been performed in the early supergravity bibliography in \cite{Jasinschi:1983wr,Jasinschi:1984cx} to yield the effective scalar potential, but the actual existence of a condensate was later questioned \cite{Odintsov:1988wz,Buchbinder:1989gi}. 
More recently, the condensation was revisited and put on firmer footing in \cite{Ellis:2013zsa,Alexandre:2013iva,Ishikawa:2019pnb}. Furthermore, both the Fierz ambiguity and the wavefunction renormalization of the condensate were addressed \cite{Alexandre:2014lla}. A comprehensive and analytic review of all this work can be found in \cite{Houston:2015ygm}. 
The crucial result in these articles pertains to the behaviour of the effective scalar potential for the condensate $\sigma(x)$, which turns out to be {\it tachyonic} around the original central vacuum. This, therefore, indicates that an uplift from an AdS vacuum to a de Sitter one with a pure non-linear realization may be inherently unstable. 
The drawback of these articles, however, is that a manifestly supersymmetric analysis is missing and a precise and controlled form of the effective theory for the condensates is elusive. 
This also means that, if one wanted to extend the analysis to a matter coupled Supergravity, for example to the K\"ahler modulus of KKLT, the full procedure would have to be performed from scratch. 

In this work we make a first step towards filling this gap by providing a manifestly supersymmetric description of the composite states. 
One could suspect that the effect that manifests itself in a supergravity framework as gravitino condensates may already exist in the rigid limit in the form of goldstino condensates. 
Indeed, such behaviour of the Volkov--Akulov fermion can be justified by the fact that the theory contains four-Fermi interactions, just like the fermions in the Nambu--Jona-Lasinio model. 
In a typical Nambu--Jona-Lasinio setup the way in which the effective theory for the bound states is uncovered is by recasting the four-Fermi interactions $\left(\overline \Psi \Psi \right)^2$ as $-\sigma^2 + \sigma \overline \Psi \Psi$, with $\sigma$ taken to be auxiliary at some UV scale, and then following the flow of the theory to the IR. 
This generates a kinetic term for the scalar $\sigma$ and also gives rise to new contributions to its effective potential, which leads to the formation of a new vacuum where the condensation takes place. 
However, for a single goldstino, a large-N expansion that helps to control the diagrams in the Nambu--Jona-Lasinio model is not available; as a consequence, a perturbative loop-diagram analysis is not tractable and a direct non-perturbative analysis is required. 
Indeed, the Nambu--Jona-Lasinio model can be discussed for a single fermion species if a non-perturbative renormalization group flow is utilized \cite{Jaeckel:2002rm}. 

Our procedure here is thus the following. We first recast the full Volkov--Akulov model in terms of two unconstrained superfields, $X$ and $T$, where the latter is a Lagrange multiplier that is integrated out to impose the nilpotency condition $X^2=0$ on the former \cite{Rocek:1978nb,Casalbuoni:1988xh,Komargodski:2009rz}. 
Once the nilpotency is imposed, we recover the typical Volkov--Akulov model. 
To uncover the low energy description of the theory, and possibly the existence of light composite states, we have to track the flow of the theory towards the IR. 
We do this by following the flow determined by the {\it exact renormalization group} (ERG) equations within a {\it supersymmetric} rendition of the local potential approximation (which we will denote as SLPA) \cite{Polchinski:1983gv,Zumbach:1994vg,Zumbach:1994kc,Harvey-Fros:1999qpe,Morris:1994ie,Ball:1993zy,Litim:2018pxe}. 
In particular, we perform this analysis preserving supersymmetry off-shell and we keep track of the full flow of the K\"ahler potential. 
Our final result verifies the emergence of composite states, which now fall into standard chiral multiplets with {\it linearly realized} supersymmetry, with an effective low energy description where the Volkov--Akulov K\"ahler potential $K_\text{VA} = X \overline X$ is replaced by
\be
K_\text{composite states} = Z_X \, X \overline X + Z_T \, T \overline T + \text{higher order terms} \,, 
\ee
where the $Z's$ indicate the wavefunction renormalizations, and where the superpotential remains the same as in the original Volkov--Akulov model, 
\be
W = f X + \frac12 T X^2 \,. 
\ee
A study of the potential of the resulting supersymmetric theory shows that tachyons are generated near the origin of the ($X$,$T$) field space, signaling an inherent non-perturbative instability of the pure Volkov--Akulov model. 
The compositeness of $T$ is manifested by the vanishing $Z_T$ at the UV point of the flow\footnote{Our approach can be described as replacing a Lagrangian where supersymmetry is non-linearly realized by an equivalent low energy model where supersymmetry has, instead, a linear realization. We could therefore say that we have an emergence of supersymmetry in the IR. Although the concept of ``emergent supersymmetry'' has previously appeared in \cite{Fei:2016sgs,Gies:2017tod}, one should investigate whether our findings could be embedded in that framework.}.

A further step would be to perform a similar procedure in Supergravity using again an ERG flow \cite{Granda:1997xk,Percacci:2013ii}. An important obstacle to performing this analysis in Supergravity is the requirement of a supersymmetric regulator.
However, for small fields, one can trust the supersymmetric analysis: the supergravity effects should only enter as we probe larger distances in field space, where the $1/M_P$ effects will become important. 
Therefore, the same low energy effective description for the Volkov--Akulov composite states can be justified to hold also in Supergravity, but only near the field space origin. 
We use this approximation to show that the Volkov--Akulov model, now coupled to 4D $N=1$ Supergravity as in \cite{Bergshoeff:2015tra}, again suffers from tachyonic instabilities; this verifies the earlier results regarding tachyons due to gravitino condensation \cite{Jasinschi:1984cx,Alexandre:2014lla}. 
The same holds also for the KKLT supergravity embedding, when we introduce a single K\"ahler modulus \cite{Ferrara:2014kva}. 
Our results actually indicate that, if a 4D $N=1$ supergravity theory flows in the IR to the model of \cite{Bergshoeff:2015tra}, then it will suffer from the same instability (as it happens for the KKLT 4D $N=1$ embedding of \cite{Ferrara:2014kva}). 
The only way to avoid the instability associated with the formation of these composite states would be to always have some additional light states (possibly of non-perturbative origin) surviving in the IR and either disrupt the procedure that we described or alter the flow.

\section{Composite supersymmetry from the Volkov--Akulov model}

\subsection{The setup}

The four-dimensional Volkov--Akulov model \cite{Volkov:1973ix} can be described in terms of a constrained superfield $X$ that satisfies \cite{Rocek:1978nb,Casalbuoni:1988xh} 
\be
\label{X-nil}
X^2 = 0 \,, \quad X| = \frac{G^2}{2F^X} \,, 
\ee
with the Lagrangian, defined in terms of a K\"ahler potential $K$ and a superpotential $W$, 
\be
{\cal L} = \int d^4 \theta \, K + \left(\int d^2\theta \, W + \text{c.c.} \right) = \int d^4 \theta |X|^2 + \left(\int d^2\theta f X + \text{c.c.} \right) \,. 
\ee
This procedure is further described in \cite{Komargodski:2009rz} and the component form of the Lagrangian is 
\be
\label{VA-VA}
{\cal L} = 
- f^2 
+ i G \sigma^m  \partial_m \overline G 
- \frac{1}{4 f^2} \overline G^2 \partial^2 G^2 
- \frac{1}{16 f^6} G^2 \overline G^2 \partial^2 G^2 \partial^2 \overline G^2 \, 
\ee
(using the mostly-minus convention for the spacetime metric, $\eta_{mn}=\text{diag}(+1,-1,-1,-1)$). 
The presence of the non-linear self-interaction terms opens the possibility of composite states of two goldstini, e.g. $G^2/f$. We wish to investigate the possible description of such states in the low energy theory\footnote{Note that, classically, the various forms of the Volkov--Akulov model have been shown to be equivalent \cite{Kuzenko:2011tj,Cribiori:2016hdz}, and always reduce to \eqref{VA-VA} in the component form.}.

First, we bring the theory into a form where supersymmetry is linearly realized, and use it as the form of the theory at the UV point. 
To this end we introduce a Lagrange multiplier multiplet $T$ (associated with no term in the K\"ahler potential) by changing the superpotential to 
\be
\label{SUPERUV}
W = f X + \frac12 T X^2 \,. 
\ee
By varying $T$ we recover the superspace condition \eqref{X-nil}. Note that the Volkov--Akulov model naturally comes with a UV scale given by the supersymmetry breaking scale, which is controlled by $f$; it is therefore natural, though not mandatory, to match the two descriptions at that energy scale.

At this point, let us prove that \eqref{SUPERUV} is the most general form of the superpotential that one can write in the UV point and that $K=|X|^2$ is the most general K\"ahler potential.

We firstly observe that, by definition, the K\"ahler potential has no dependence on $T$ in the UV point. We then assume that the superpotential takes the form $W = f X + \frac12 T X^2 + P(X)$, where $P(X) = \sum_{n \geq 0} P_n X^n = P_0 + P_1 X + P_2 X^2 + P_3 X^3 + \dots$ is some arbitrary analytic function (as it can always taken to be, since we are dealing with a superpotential), and we further note that we can not have terms like $1/X^p$ with $p>0$, because they will become ill-defined once the nilpotency condition is imposed. Now, $P_0$ clearly drops out due to the $\int d^2 \theta$ and $P_1$ can be absorbed into $f$. So, we remain with $f X + \frac12 T X^2 + X^2 \tilde P(X)$, where $\tilde P(X) = P_2 + P_3 X + P_4 X^2 + \dots$, which is still granted to be an analytic function of $X$. If we simply shift $T$ to $T - 2 \tilde P(X)$, we are left with our original superpotential \eqref{SUPERUV}. This shift is consistent precisely because the function $\tilde P(X)$ is analytic. Moreover, there are no $T$-dependent terms in the K\"ahler potential, so this shift does not generate new terms.\\
Analogously, let us consider the K\"ahler potential $K = |X|^2 + \sum_{m,n \geq 0} M_{nm} X^n \overline X^m$ assuming by {\it fiat} that it can be expanded in a power series (as part of our general assumptions). 
We can directly see that $M_{00}$, $M_{01}$ and $M_{10}$ drop out due to the $\int d^4 \theta$, whereas $M_{11}$ is absorbed by redefining the $|X|^2$ term. Then, we are left with $K = |X|^2 + X^2 \tilde M(X, \overline X) + \overline X^2 \left(\tilde M(X, \overline X)\right)^*$, where $\tilde M(X, \overline X) = \sum_{m,n \geq 0} \tilde M_{nm} X^n \overline X^m$. Now, once simply shifting $T$ to $T + \frac12 \overline D^2 \tilde M(X, \overline X)$, we are left with our original K\"ahler potential.\\
We are thus working with the most general K\"ahler potential and superpotential at the UV point. 

The second step of our procedure is to lower the energy scale at which we probe the theory via a renormalization group (RG) flow and this will make the superfield $T$ acquire a kinetic term. 
The interpretation of this is that the new standard chiral superfields $X$ and $T$ now describe composite states of the original fermion goldstino. 
This means that the partition function of the system has the form 
\be
{\cal Z} = \frac{1}{N_0} \int {\cal D}[T] {\cal D}[X] \, e^{i S[T,X]} \,. 
\ee
Notice that, once we start integrating out individual components of the superfields, up to the overall normalization, the partition function reduces to $\int {\cal D}[G]\,e^{i S[G]}$, where $S[G]$ comes from \eqref{VA-VA}. We therefore get back the partition function for the Volkov--Akulov model.  

We will study the RG flow by using the exact renormalization group method discussed in \cite{Polchinski:1983gv,Ball:1993zy,Litim:2018pxe}.
To follow this method it is typical to introduce all possible terms in the K\"ahler potential and the superpotential that are consistent with the symmetries that are expected to remain preserved, such that the flow can be properly described. Some terms may stay trivially zero throughout the flow depending on the initial conditions and the subsequent flow. Since such a procedure actually requires to introduce {\it infinite} terms, in order to make it tractable the local potential approximation (LPA) has been devised (see e.g. \cite{Zumbach:1994vg,Zumbach:1994kc,Harvey-Fros:1999qpe,Morris:1994ie}). 
The idea standing behind the LPA is that it may be advantageous to apply approximations directly at the level of a differential equation rather than applying approximations at the level of the solution. 
In this way the LPA simply works by truncating all the contributions of derivative interactions to the ERG equations, which are expected to be irrelevant to the IR dynamics, anyway. 
For a supersymmetric theory, however, since the supersymmetry transformations contain derivatives, different orders of derivative interactions mix: a crude LPA would then explicitly violate supersymmetry. Here, we will then use a type of local potential approximation that allows us to preserve supersymmetry manifestly, which we will simply call, as already mentioned in the introduction, the {\it supersymmetric} local potential approximation (SLPA). 
It is easier to describe this approximation directly in the superspace language\footnote{A superspace account of the RG flow has been presented in \cite{Rosten:2008ih}, but for our purposes we will ultimately work with the component form.}.
First of all, supersymmetric theories with chiral superfields ${\cal X}^i$ are in any case described by two types of superspace integrals, which are $\int d^4 \theta$ and $\int d^2 \theta$. If we think of the functions that we can insert in those integrals as having a (superspace) derivative expansion, then we can have 
\be
\label{DER-EXP}
\int d^4 \theta \left( K({\cal X}^i,\overline{\cal X}^j) + {\cal O} \left(\partial_m, D_\alpha, \overline D_{\dot \beta}\right) \right)
+ \left( \int d^2 \theta \, W({\cal X}^i) + \text{c.c.} \right) \,,
\ee
where the first term corresponds to the K\"ahler potential and the second one to the superpotential. 
Our SLPA can be now simply defined by stating that we will not keep track of superspace higher derivative terms. 
This means that any term in \eqref{DER-EXP} that is not a a part of $K$ or $W$ will always be ignored. 
Similarly, since we will be working in component form, we will not keep track of any terms that do not correspond to a K\"ahler potential or a superpotential. 
We will ignore these terms when they are generated during the flow and we will not take into account their backreaction into the ERG equations. 

Even though these types of approximations are in standard use in the ERG literature, it is important to understand and be conscious of their limitations regarding the physical conclusions that one can draw. We will discuss these issues carefully in the next section. 
For now, we proceed within this approximation and derive the SLPA RG flow of our model. We will, however, note immediately that in our chosen approach, even though we will not be able to see the anomalous dimensions of the fields, we will have a certain handle on the wavefunction renormalizations because of the presence of other fields besides the physical scalars. 

Even within the LPA, one still makes an expansion in the fields to derive the RG flow, which means that we still need to make an expansion in terms of superfields in the K\"ahler potential. 
In our case, we will make an educated guess and keep only the terms that provide a self-consistent flow. 
Indeed, the gratifying result of applying the SLPA is that we get a solution to the ERG in a closed form that includes only a handful of new terms in the K\"ahler potential. 
As we will see, it will suffice to work with the K\"ahler potential
\be
\label{K-total}
K = \alpha |X|^2 + \beta |T|^2 + g \, |T|^2 |X|^2 + \frac14 q \, |X|^4 \, 
\ee
and the superpotential 
\be
\label{SW}
W = f X + \frac12 T X^2 \,, 
\ee
where we have both dimensionful and dimensionless couplings: $[\alpha] = 0$, $[\beta] = 0$, $[g] = -2$ and $[q] = -2$. 
This theory is defined with a UV cut-off $\Lambda$. 
We want to study its properties as we integrate out the high energy modes down to a lower scale $\mu$, with 
\be
\mu \leq \Lambda \,. 
\ee
This is often called the renormalization scale and it is related to the energy scale at which we probe the theory. 
Then, the renormalization time is defined as 
\be
t = \log \frac{\Lambda}{\mu} \,, 
\ee
and the flow to the IR is described by $\mu \downarrow$ and $t \uparrow$. 
To match with the Volkov--Akulov action at the UV point, that is when $\mu = \Lambda$, we want to find the RG flow of the couplings with the boundary conditions 
\be
\label{boundary}
\alpha\Big{|}_{\mu = \Lambda} = 1 \,, \quad 
\beta\Big{|}_{\mu = \Lambda} = 0 \,, \quad 
g\Big{|}_{\mu = \Lambda} = 0 \,, \quad 
q\Big{|}_{\mu = \Lambda} = 0 \,. 
\ee  
Indeed, we notice that $\beta(\mu = \Lambda) = 0$ and $g(\mu = \Lambda) = 0$ are required in order for the superfield $T$ to act as a Lagrange multiplier at the UV scale and $q(\mu=\Lambda) = 0$ because, as we have showed above, $K = |X|^2$ is the most general expression of the UV K\"ahler potential that we can have.

Following \cite{Polchinski:1983gv,Ball:1993zy,Litim:2018pxe}, we re-organize the action into propagator and interaction parts. 
We do this in a supersymmetric manner by appropriately splitting the K\"ahler potential.
In particular, we have 
\be
\label{K-prop}
K_{\text{prop.}} = c^{-1} |X|^2 + c^{-1} |T|^2  \,, 
\ee
where $c$ is a scale-dependent regularization function, which we will discuss momentarily, with $[c]=0$, and 
\be
K_{\text{int.}} = (\alpha - 1) |X|^2 + (\beta - 1) |T|^2 + g |T|^2 |X|^2 + \frac14 q |X|^4 \,. 
\ee
Note that we leave the background field dependence in the interaction part of the K\"ahler potential. 
Similarly, the superpotential naturally only contributes to the interactions: 
\be
W_{\text{int.}} = W \,. 
\ee 
Moreover, since no background-independent mass term exists neither for $X$ nor for $T$ in the UV, and it will not be generated during the flow\footnote{This is true at least perturbatively, and can also be checked explicitly within our SLPA analysis.}, we do not include a mass for these fields in the propagator piece \eqref{K-prop}. 

Returning to the propagator regularization function $c$, working in Euclidean momentum space, we can write
\be
\label{cc11} 
c = \sum_{n=0}^{+\infty} c_n \hat p^{2n} 
\quad \text{with} \quad \hat p = \mu^{-1} p \,, 
\ee
as long as some basic asymptotic properties are satisfied \cite{Litim:2018pxe}, where now we have $|\hat p| \leq 1$.\\ 
For our calculations we will not need to work with an explicit form for $c$, but for the benefit of the reader we can give as a simple example the expression $c(p,\mu) = \left( 1 - \hat p^2 \right) \Theta\left( 1 - \hat p^2 \right)$, which is discussed in \cite{Litim:2018pxe}. Such a regulator is manifestly supersymmetric, because the component field propagator terms are described collectively by the superspace integral $\int d^4 \theta \, (c(-\partial^2))^{-1} \left(|X|^2 + |T|^2 \right)$ and the superspace derivatives (the $D's$) commute with any combination of spacetime derivatives.\\ 
In addition, since we will only be interested in the vacuum (in)stability of the theory, we will not include any source terms in our analysis or any derivative interactions. As discussed in \cite{Ball:1993zy}, this allows, together with the use of the $c$ function \eqref{cc11} in the propagator piece, to simplify the calculations. We importantly stress that the choice of regulator does not ultimately affect any physical results in the IR.
We further define the renormalization time derivative of the regulator $c$ as
\be \label{C2}
\dot c \equiv \frac{\partial}{\partial t}c 
= -\mu \partial_\mu c 
= p \partial_p c 
= \hat p \partial_{\hat p} c 
=  2 \hat p^2 c_1 + {\cal O}(\hat p^4) \,. 
\ee  

In our procedure we will rely heavily on supersymmetry and, because we will work directly in component form, this means that, within the SLPA, we will only evaluate the flow of the coefficients of the auxiliary field potential that are related to the K\"ahler potential. In other words, we will only keep track of the terms of the form 
\be
{\cal L} = g_{i \overline j} F^i \overline F^j + \dots \,, 
\ee 
and from these terms we will deduce the full flow of the K\"ahler potential. A similar procedure is used in \cite{Brignole:2000kg} for the evaluation of the one-loop K\"ahler potential. 
For the study of the flow we will be using the Euclidean conventions of \cite{Peskin:1995ev}, where the Lorentzian conventions correspond to $\eta_{mn} = \text{diag}(+1,-1,-1,-1)$, which also matches with the conventions of \cite{Polchinski:1983gv}. 

We now go to components and work with the momentum space fields. 
We define 
\be
\hat X(\hat p) = \frac{1}{\sqrt 2} (\phi + i \chi) \,, \quad  \hat T(\hat p) = \frac{1}{\sqrt 2} (\tau + i \sigma) \,, 
\ee
and 
\be  \label{auxnorm}
\hat F^X(\hat p) = F_1 + i F_2 \,, \quad \hat F^T(\hat p) = B_1 + i B_2 \,, 
\ee
which have vanishing mass dimensions $[\phi] = 0$, $[F_1] = 0$, etc. 
We further follow \cite{Polchinski:1983gv} and obtain the propagator part of the action\footnote{To get to Euclidean momentum space, 
we first Wick-rotate and then we go to momentum space.}  
\be
\begin{aligned}
L_{\text{prop.}} = & \int \frac{d^4 \hat p}{(2 \pi)^4} \left[ - \frac12 \hat p^2 c^{-1} ( \phi^2  + \chi^2 + \tau^2 + \sigma^2 ) \right] +
\\
& +  \int \frac{d^4 \hat p}{(2 \pi)^4} \left[ c^{-1} ( F_1^2  + F_2^2 + B_1^2 + B_2^2 ) \right] + 
\\
& + \text{fermion propagator terms} \,, 
\end{aligned}
\ee
where we have $\phi^2 = \phi(\hat p) \phi(- \hat p)$, $F_1^2 = F_1(\hat p) F_1(-\hat p)$, etc. 
We also define dimensionless couplings via
\be \label{zgzg}
\gamma = \mu^2 g(\mu) \quad \text{and} \quad \zeta = \mu^2 q(\mu) \,,
\ee
so that $[\gamma] = 0$ and $[\zeta] = 0$. 
For the interacting part of the component field action we have 
\be
\begin{aligned}
L_{\text{int.}} = \ &  \frac12 B_1 (\phi^2 - \chi^2) + F_1 (\phi \tau - \chi \sigma) + (\alpha-1) F_1^2 + (\beta-1) B_1^2 \ +
\\
& + \frac{\gamma}{2} \left[ F_1^2 (\tau^2 + \sigma^2) + B_1^2 (\phi^2 + \chi^2) \right] + \frac{\zeta}{2} F_1^2 (\phi^2 + \chi^2) + \dots \,, 
\end{aligned}
\ee
where the dots stand for many other interaction terms and many terms including fermions. 
Here $L_{\text{int.}}$ is only a formal compact expression and it really means that we should treat all terms in the expansion in the form 
\be \label{NOT}
L_{\text{int.}} = \int \frac{d^4 \hat p_1 \dots d^4 \hat p_n}{(2 \pi)^{4n-4}} \ \hat Y_{A_1 \dots A_n}(t) 
\ \hat \Psi_{A_1}(\hat p_1) \dots \hat \Psi_{A_n}(\hat p_n)  \ \delta \left(\sum_{i=1}^n \hat p_i \right) \,, 
\ee
with $[\hat Y_{A_1 \dots A_n}(t)] = 0$ and $[\hat \Psi_{A_i}(\hat p_i)] = 0$ (for any $A_i$). 
A sample of illustrative terms gives 
\be \label{LINT}
\begin{aligned}
L_{\text{int.}} = & \int \frac{d^4 \hat p_1 d^4 \hat p_2}{(2 \pi)^{4}} \ (\alpha(t) - 1) \ F_1(\hat p_1) F_1(\hat p_2) \ \delta(\hat p_1 + \hat p_2) \ +
\\
& + \int \frac{d^4 \hat p_1 d^4 \hat p_2 d^4 \hat p_3 d^4 \hat p_4}{(2 \pi)^{12}} \frac{\gamma(t)}{2} F_1(\hat p_1) F_1(\hat p_2) \tau(\hat p_3) \tau(\hat p_4) \
\delta(\hat p_1 + \hat p_2 + \hat p_3 + \hat p_4) \ +
\\
& + \dots \,.
\end{aligned}
\ee
For completeness, let us mention that the full Euclidean partition function is 
\be
{\cal Z} = \int {\cal D}[X] {\cal D}[T] \, e^{ L_{\text{prop.}} + L_{\text{int.}} } \,,
\ee
just as in \cite{Polchinski:1983gv}. 
As already mentioned, the symbol $L$ clearly refers to an action, but we keep the notation of \cite{Polchinski:1983gv}, using the symbol $L$ instead of $S$. Note also, once again, that all fields, momenta and couplings are dimensionless. 

We can now use the ERG equation from \cite{Polchinski:1983gv} to obtain
\be
\label{PEQ}
\begin{aligned}
\dot L_{\text{int.}} = & -  \int d^4 \hat p \ \frac{(2 \pi)^4}2 \ \hat p^{-2} \dot c(\hat{p}) \sum_{\varphi^a = (\phi, \chi, \tau, \sigma)}  
\left( \frac{\partial L_{\text{int.}}}{\partial \varphi^a} \frac{\partial L_{\text{int.}}}{\partial \varphi^a} 
+ \frac{\partial^2 L_{\text{int.}}}{\partial \varphi^a \partial \varphi^a} \right) +
\\ 
& + \int d^4 \hat p \ (2 \pi)^4 \ \dot c(\hat{p}) \sum_{h^a = (F_1, F_2, B_1, B_2)}  
\left( \frac{\partial L_{\text{int.}}}{\partial h^a} \frac{\partial L_{\text{int.}}}{\partial h^a} 
+ \frac{\partial^2 L_{\text{int.}}}{\partial h^a \partial h^a} \right) +
\\ 
& + \text{fermion propagator terms} \,.  
\end{aligned} 
\ee
We have independent sums in the ERG equation \eqref{PEQ} because all our scalars have diagonal propagator terms. 
Notice that in a slight abuse of notation, we denote by partial derivatives what should really be understood as variational derivatives, with a momentum matching $\delta$-function, i.e., say,
\be
\frac{\partial \varphi^a(p)}{\partial \varphi^b(k)} = \delta_b^a \delta^{(4)}(p-k) \,, 
\ee
The lack of a $(2 \pi)^4$ factor on the $\delta$-function above is due to the fact that it is already included explicitly in the expression \eqref{PEQ}. This choice of notation and normalization, which will be used throughout the rest of this section, corresponds to the one in \cite{Polchinski:1983gv}.

At this stage we have to insert the interacting action \eqref{LINT} into the ERG equation and equate term by term to deduce the flow equations. 
We stress that one only needs to look at the terms related to the auxiliary field potential, namely 
\be
F_1^2 \,, \quad  \tau^2  F_1^2 \,, \quad \text{etc.} \,.
\ee
To this end the third line in \eqref{PEQ} does not play any role as one can check by considering the fermionic contribution to the ERG (see e.g. \cite{Ball:1993zy}) and the fermion couplings to the auxiliary fields, which are only linear in the auxiliary fields (see e.g. \cite{Wess:1992cp}). 
Therefore, we will focus on the first two lines of \eqref{PEQ} and we will only use the fermions as a cross-check. 

One can easily prove that the superpotential does not receive any corrections within the SLPA by checking that no terms linear in the auxiliary fields are generated. This can be seen faster by writing \eqref{PEQ} in a form where the scalars and the auxiliary fields are recast to be complex by a simple chain rule.

\subsection{The RG flow within the supersymmetric local potential approximation} \label{sec22}

Our aim is to find the flow equations for the couplings $\alpha$, $\beta$, $\zeta$ and $\gamma$. 

Let us first look at the equation governing the flow of $\zeta$. 
This means that on the left hand side of \eqref{PEQ} we want to focus on the term
\be
\frac12 (\dot \zeta + 2 \zeta) F_1^2 \phi^2 
= \int \frac{d^4 \hat p_1 d^4 \hat p_2 d^4 \hat p_3 d^4 \hat p_4}{(2 \pi)^{12}} \frac12 (\dot \zeta + 2 \zeta) F_1(\hat p_1) F_1(\hat p_2) \phi(\hat p_3) \phi(\hat p_4) \delta\left(\sum_{i=1}^4 \hat p_i\right) \,. 
\ee 
On the right hand side we have the term 
\be \label{F2}
- \int d^4 \hat k \frac{(2 \pi)^4}2 \hat k^{-2} \dot c(\hat k)  
\frac{\partial (F_1 \phi \tau)}{\partial \tau(\hat k)} \frac{\partial (F_1 \phi \tau)}{\partial \tau(-\hat k)}  \,, 
\ee 
and a variety of seemingly relevant terms related to the auxiliary field propagators as, for instance, the term 
\be \label{F4}
+ \int d^4 \hat k (2 \pi)^4 \dot c(\hat k) 
\left( \frac{\partial \left( (\alpha-1) F_1^2 \right)}{\partial F_1(\hat k)} \frac{\partial \left( \frac{\zeta}{2}  F_1^2 \phi^2 \right)}{\partial F_1(-\hat k)} \right) \,.  
\ee 
As we will now see, only the term \eqref{F2} actually contributes to this part of the flow, whereas \eqref{F4} contributes to derivative interactions. 
Indeed, up to an overall coefficient, \eqref{F4} gives  
\be 
(\alpha-1) \zeta \, 
\int d^4 \hat p_1 d^4 \hat p_2 d^4 \hat p_3 d^4 \hat p_4 \, 
\dot c(\hat p_1) \, F_1(\hat p_1)  F_1(\hat p_2) \phi(\hat p_3) \phi(\hat p_4) \, \delta\left(\sum_i \hat p_i\right) \,, 
\ee
which means that this term contributes only to derivative terms: in fact, from \eqref{C2} we have 
\be
\int d^4 \hat p_1 \, \dot c(\hat p_1) F_1(\hat p_1) \sim \int d^4 \hat p_1 \left( 2 \hat p_1^2 c_1 + {\cal O}(\hat p_1^4) \right) F_1(\hat p_1) \,. 
\ee 
In a similar way, we can see that the only relevant part of \eqref{F2} is given by the $c_1$ part of the expansion 
\be
\frac12 \hat k^{-2} \dot c(\hat k)  = c_1 + {\cal O}(\hat k^2) \,.
\ee
We then evaluate 
\be
\begin{aligned}
\frac{\partial (F_1 \phi \tau)}{\partial \tau(\hat k)} 
& = \int \frac{d^4 \hat p_1 d^4 \hat p_2 d^4 \hat p_3}{(2 \pi)^{8}} F_1(\hat p_1) \phi(\hat p_2) \delta(\hat p_3 - \hat k) \delta(\hat p_1 + \hat p_2 + \hat p_3) =
\\
& = \int \frac{d^4 \hat p_1 d^4 \hat p_2}{(2 \pi)^{8}} F_1(\hat p_1) \phi(\hat p_2) \delta(\hat p_1 + \hat p_2 + \hat k) \,, 
\end{aligned}
\ee 
and 
\be
\frac{\partial (F_1 \phi \tau)}{\partial \tau(- \hat k)} = \int \frac{d^4 \hat p_3 d^4 \hat p_4}{(2 \pi)^{8}} F_1(\hat p_3) \phi(\hat p_4) \delta(\hat p_3 + \hat p_4 - \hat k)  \,. 
\ee
Finally, \eqref{F2} becomes 
\be 
\begin{aligned}
- c_1 (2 \pi)^4 \int d^4 \hat k \frac{\partial (F_1 \phi \tau)}{\partial \tau(\hat k)} \frac{\partial (F_1 \phi \tau)}{\partial \tau(-\hat k)} &=\\& =-c_1 \int \frac{\prod_{i=1}^{4} d^4 \hat p_i }{(2 \pi)^{12}} F_1(\hat p_1) \phi(\hat p_2) F_1(\hat p_3) \phi(\hat p_4) \delta\left(\sum_{i=1}^{4} \hat p_i\right) \,.
\end{aligned}
\ee
This means that in the compact notation \eqref{NOT} the relevant part of \eqref{PEQ} takes the form $\frac12 (\dot \zeta + 2 \zeta) F_1^2 \phi^2 = - c_1 F_1^2 \phi^2$, which delivers 
\be
\label{zetad}
\dot \zeta = - 2 \zeta - 2 c_1 \,, 
\ee
where the $-2 \zeta$ is due to the fact that $\zeta$ originates from a dimensionful coupling.\\ 
Similarly, for the coupling $\gamma$ we are bound to get 
\be
\label{gammad}
\dot \gamma = - 2 \gamma - 2 c_1 \,. 
\ee
In this way we see that the tree level interactions from the superpotential contribute to the flow of the higher order terms in the K\"ahler potential. 

We will now analyze the equations that govern the flow of $\beta$. 
The relevant term of left hand side of \eqref{PEQ} is
\be
\dot \beta  B_1^2 = \int \frac{d^4 \hat p_1 d^4 \hat p_2}{(2 \pi)^{4}} \dot \beta  B_1(\hat p_1)  B_1(\hat p_2) \delta(\hat p_1 + \hat p_2) \,,
\ee 
and on the right hand side we have 
\be
\label{SCB}
- \int d^4 \hat k \frac{(2 \pi)^4}2 \hat k^{-2} \dot c(\hat{k}) \sum_{\varphi^a = (\phi, \chi, \tau, \sigma)} \frac{\partial^2 \left( \frac{\gamma}{2}  B_1^2 (\phi^2 + \chi^2)  \right)}{\partial \varphi^a(\hat k) \partial \varphi^a (-\hat k)} \,, 
\ee
and a variety of seemingly relevant terms related to the auxiliary field propagators, as the term 
\be
\label{AUB}
+ \int d^4 \hat k (2 \pi)^4 \dot c(\hat{k}) \sum_{h^a = (F_1, F_2, B_1, B_2)} \left( \frac{\partial \left( (\beta-1) B_1^2 \right)}{\partial h^a(\hat k)} \frac{\partial \left( (\beta-1) B_1^2 \right)}{\partial h^a(-\hat k)} \right) \,. 
\ee
As we will see right away, the term \eqref{AUB} and other similar terms from the second line of \eqref{PEQ} do not enter this part of the flow and can be safely ignored. 
Indeed, focusing on \eqref{AUB} we find 
\be
\begin{aligned}
& \int d^4 \hat k (2 \pi)^4 \dot c(\hat k) \sum_{h^a = (F_1, F_2, B_1, B_2)} \left( \frac{\partial \left( (\beta-1) B_1^2 \right)}{\partial h^a(\hat k)} \frac{\partial \left( (\beta-1) B_1^2 \right)}{\partial h^a(-\hat k)} \right) =
\\
& = 4 (\beta-1)^2 \int d^4 \hat k (2 \pi)^4 \dot c (\hat k) 
\left( \int \frac{d^4 \hat p_1 d^4 \hat p_2}{(2 \pi)^{4}} 
\ B_1(\hat p_1) \delta(\hat p_2 - \hat k)
 \delta(\hat p_1 + \hat p_2)  \right) \times
\\ 
&\qquad \qquad \qquad \qquad \qquad \qquad \times \left( \int \frac{d^4 \hat p_3 d^4 \hat p_4}{(2 \pi)^{4}} 
\ B_1(\hat p_3) \delta(\hat p_4 + \hat k)
 \delta(\hat p_3 + \hat p_4)  \right) =
\\
& = 4 (\beta-1)^2 \int \frac{d^4 \hat p_1 d^4 \hat p_2 d^4 \hat p_3 d^4 \hat p_4}{(2 \pi)^{4}} \dot c (\hat p_2) B_1(\hat p_1) 
 \delta(\hat p_1 + \hat p_2)  B_1(\hat p_3)   \delta(\hat p_4 + \hat p_2)  \delta(\hat p_3 + \hat p_4) =
\\
& = 4 (\beta-1)^2 \int \frac{d^4 \hat p_1 d^4 \hat p_2}{(2 \pi)^{4}} \dot c (\hat p_2)  B_1(\hat p_1)  B_1(\hat p_2)  \delta(\hat p_1 + \hat p_2) \,. 
\end{aligned}
\ee 
From \eqref{C2} we see that here effectively $\dot c(\hat p_2) = 2 c_1 \hat p_2^2 + {\cal O}(\hat p_2^4)$, which means that \eqref{AUB} does not contribute to the flow of $\beta$; it contributes, instead, to the higher order derivative terms. 
Therefore, the flow of $\beta$ is controlled by \eqref{SCB}.\\ 
We have two terms in \eqref{SCB}, but their contribution is the same: we will work out one of the two terms and double the result.
We have 
\be
\begin{aligned}
&
- \int d^4 \hat k \frac{(2 \pi)^4}2 \hat k^{-2} \dot c(\hat k)  \frac{\gamma}{2} \frac{\partial^2 \left( B_1^2 \phi^2 \right)
}{\partial \phi(\hat k) \partial \phi(-\hat k)} =
\\ 
& = - \frac{(2 \pi)^4 \gamma}2 \int d^4 \hat k  \hat k^{-2} \dot c(\hat k) \int \frac{d^4 \hat p_1 d^4 \hat p_2 d^4 \hat p_3 d^4 \hat p_4}{(2 \pi)^{12}} B_1(\hat p_1) B_1(\hat p_2) 
\delta(\hat p_3 - \hat k) \delta(\hat p_4 + \hat k) \delta \left(\sum_{i=1}^4 \hat p_i \right) =
\\ 
&  = - \frac{\gamma}{2} \int \frac{d^4 \hat p_1 d^4 \hat p_2 d^4 \hat p_3 d^4 \hat p_4}{(2 \pi)^{8}} \ \hat p_3^{-2} \dot c(\hat p_3) B_1(\hat p_1) B_1(\hat p_2) \delta(\hat p_3 + \hat p_4) \delta(\hat p_1 + \hat p_2 + \hat p_3 + \hat p_4) =
\\ 
& = - \frac{\gamma}{2} \int \frac{d^4 \hat p_1 d^4 \hat p_2 d^4 \hat p_3}{(2 \pi)^{8}} \ \hat p_3^{-2} \dot c(\hat p_3) 
B_1(\hat p_1) B_1(\hat p_2) \delta(\hat p_1 + \hat p_2)
 \,. 
\end{aligned}
\ee
Now, we postulate that the momentum integral over $p_3$ takes a value $N$, which is regulator-dependent, and is given by 
\be
\label{NNN}
N = \frac{1}{2} \int \frac{d^4 \hat p_3 }{(2 \pi)^4} \hat p_3^{-2} \dot c(\hat p_3) \,. 
\ee 
The exact value of $N$ can thus be evaluated only once we have a specific regularization scheme at hand. 
We conclude that 
\be
- \int d^4 \hat k \frac{(2 \pi)^4}2 \hat k^{-2} \dot c(\hat{k}) \frac{\gamma}{2} \frac{\partial^2 \left(B_1^2 \phi^2 \right)
}{\partial \phi(\hat k) \partial \phi(-\hat k)} = - N \gamma 
\int \frac{d^4 \hat p_1 d^4 \hat p_2}{(2 \pi)^{4}} B_1(\hat p_1)  B_1(\hat p_2) \delta(\hat p_1 + \hat p_2) \,, 
\ee 
and, once we sum over both scalars and referring to the compact notation of \eqref{NOT}, we find that the relevant part of \eqref{PEQ} takes the form $\dot \beta  B_1^2 = 2 \times (- N \gamma)  B_1^2$, which delivers 
\be \label{betad}
\dot \beta  = - 2 N \gamma \,. 
\ee
A similar analysis for $F_1^2$ is bound to give the flow equation for the coupling $\alpha$, which is
\be
\label{alphad}
\dot \alpha  = - 2 N (\gamma + \zeta) \,. 
\ee
This completes the analysis of \eqref{PEQ} and the reader can check that no other terms are required for the self-consistent flow of the auxiliary field scalar potential. This means that the K\"ahler potential will not need higher order terms and that our flow is exact.

We directly solve the flow equations \eqref{zetad}, \eqref{gammad}, \eqref{betad} and \eqref{alphad}, together with the boundary conditions 
\be
\alpha\Big{|}_{t = 0} = 1 \,, \quad 
\beta\Big{|}_{t = 0} = 0 \,, \quad 
\gamma\Big{|}_{t = 0} = 0 \,, \quad 
\zeta\Big{|}_{t = 0} = 0 \, 
\ee
to find 
\be
\zeta = - c_1 \left( 1 - e^{-2t} \right)   \,, \quad 
\gamma = - c_1 \left( 1 - e^{-2t} \right) \, 
\ee
and 
\be 
\alpha = 1 - 2 c_1 N + 4 c_1 N \left(t + \frac12 e^{-2 t} \right) \,, \quad 
\beta = - c_1 N + 2 c_1 N \left(t + \frac12 e^{-2 t} \right)   \,. 
\ee
We conclude that the quantum effects make the Lagrange multiplier superfield $T$ become propagating and a non-ghost kinetic term requires $N c_1 > 0$, which is in accordance with the typical properties \eqref{TYPN}. 
Indeed, it is typical to have a regularization scheme where \cite{Litim:2018pxe} 
\be \label{TYPN}
c_1 < 0 \,, \quad N < 0 \,, \quad \text{so that} \quad N c_1 > 0 \,.
\ee
For example, these conditions are satisfied for the probe regulator $c(\hat{p}) = \left( 1 - \hat p^2\right) \Theta\left( 1 - \hat p^2\right)$, which gives $c_1 = -1$ and $N = - \frac{1}{32 \pi^2}$. Note that the condition $N c_1 >0$ also ensures that the multiplet $X$ doesn't become of ghost type in the IR. 
One can also work with different regulators $c(\hat p)$, and, as we have already mentioned, find similar results. 
For instance, one can use an analytic function of the form $e^{-\hat p^2 - 4 \hat p^4}$, which would give again $c_1=-1$ and $N \simeq - 10^{-3}$. An analytic regulator, however, will always have some contribution from the high UV modes because its support goes up to infinity. 

We will now shift to canonically normalized dimensionless superfields by redefining them as follows: 
\be
X \to \mu \, X / \sqrt{\alpha} \ , \quad T \to \mu \, T / \sqrt{\beta} \,,
\ee
$\mu$ being the scale compared to which we measure energies and lengths.
In addition, for the action and the superspace integrals we will have
\be
\int d^4 x \, ( ... ) \ \to \ \int d^4 x \, \mu^{-4} \, ( ... ) 
\ , \quad \int d^2 \theta \ \to \ \mu \int d^2 \theta \,, 
\ee
where the new $x$ and $\theta$ are dimensionless. 
This means that, when the redefined fields take a VEV, e.g. $\langle T \rangle = 0.1$, the original field had a VEV of the form $0.1 \times \mu / \sqrt{\beta}$.
We thus obtain a Lagrangian where all fields are dimensionless and all couplings are dressed with $\mu$, i.e. $g$ will always appear in the combination $\mu^2 g$ (and analogously for $q$). 
In the end, after we redefine the superfields $X$ and $T$ to be dimensionless and canonical, we have 
\be \label{Knorm}
\begin{aligned}
K_{\text{norm.}} = & |X|^2 + |T|^2 + \frac14 \, \frac{- c_1(1-e^{-2t})}{\left[1 - 2 c_1 N + 4 c_1 N (t + \frac{1}{2} e^{-2t}) \right]^2} |X|^4  + \\ &+ \frac{- c_1(1-e^{-2t})}{\left[1 - 2 c_1 N + 4 c_1 N (t + \frac{1}{2} e^{-2t}) \right] \left[-c_1 N + 2 c_1 N(t + \frac12 e^{-2t})\right]}  |X|^2 |T|^2 \, 
\end{aligned}
\ee
and 
\be \label{Wnorm}
\begin{aligned}
W_{\text{norm.}} = &  \frac{e^{2t} \xi_{UV}}{\left[1 - 2 c_1 N + 4 c_1 N (t + \frac{1}{2} e^{-2t}) \right]^{1/2}} X  + \\
&+ \frac12 \frac{1}{\left[1 - 2 c_1 N + 4 c_1 N (t + \frac{1}{2} e^{-2t}) \right] \left[-c_1 N + 2 c_1 N(t + \frac12 e^{-2t})\right]^{1/2}}
X^2 T \,, 
\end{aligned}
\ee
where we have defined 
\be 
f = \Lambda^2 \xi_{UV} \,. 
\ee 
Then, if we assume that the supersymmetry breaking scale of the Volkov--Akulov model serves also as the UV scale where the nilpotency of $X$ is imposed, we would have
\be
f\Big{|}_{\mu=\Lambda} = \Lambda^2 \, \text{ so that } \, \xi_{UV} = 1 \,. 
\ee
If, instead, we consider the Volkov--Akulov model to be a low energy effective description of some supersymmetric model with supersymmetry breaking scale $\sqrt{f}$, then we might wish to impose the boundary conditions \eqref{boundary} at some lower scale $\Lambda$. This is captured by choosing another value for $\xi_{UV}$, i.e. we would have
\be
f\Big{|}_{\mu=\Lambda} > \Lambda^2 \, \text{ so that } \, \xi_{UV} > 1 \,. 
\ee
We should note immediately that the qualitative results regarding the vacuum stability will not depend on $\xi_{UV}$ as long as $\xi_{UV} \geq 1$. 
In fact, the tachyonic behaviour that we will shortly demonstrate will only become more extreme as $\xi_{UV}$ increases. For this reason, we will use $\xi_{UV}=1$ in our numerical examples, as the maximally benign option.

We should also observe that, here, we consider the pure Volkov--Akulov model, which we might imagine as the low energy limit of a supersymmetry breaking model where all other degrees of freedom are sufficiently massive and can be integrated out. More generally, it could be possible that some light degrees of freedom remain in the effective theory below $\Lambda$, and could have non-trivial couplings to the nilpotent superfield $X$. In this case, one would have to include the effects of these couplings on the RG flow. However, it is worth noting that the superfield $T$ starts out as a Lagrange multiplier that does not couple to any other degrees of freedom except $X$. This means that at least for small $t$ the presence of additional light degrees of freedom in the EFT would not be able to greatly affect the evolution of $\beta$ and $\gamma$. Thus, we expect the qualitative features of the results described in the next section to remain valid even in the context of more general models.

Finally, as a non-trivial cross-check of our results, we can study the flow of the coupling $q$ with the use of the fermionic terms that are related to the relevant part of the K\"ahler potential, which are given by 
\be \label{fermi-q}
L_{int.} \! \ni \!\! \int \frac{\prod_{i=1}^4 d^4 p_{i}}{(2 \pi)^{12}}
i q \overline X(-p_1) \overline  G_{\dot \alpha}(-p_2) \overline \sigma^{m \dot \alpha \alpha } (p_3+p_4)_m G_\alpha(p_3) X(p_4) 
\delta(p_1+p_2-p_3-p_4) \! \,, 
\ee 
after Wick rotation, but still in the dimensionful notation. 
The term in \eqref{fermi-q} is influenced by the fermionic propagator and one needs the ERG equation for such fields. 
We have 
\be\label{ERGCC}
\dot L_{\text{int.}} = 
- i \int d^4 \hat k (2 \pi)^4 \hat k^{-2} \dot c(\hat{k}) \, \overline \sigma^{m \dot \alpha \alpha} \hat k_m 
\, \frac{\partial L_{\text{int.}}}{\partial \overline \lambda^{\dot \alpha}(- \hat k)} 
\frac{\partial L_{\text{int.}}}{\partial \lambda^\alpha (\hat k)} + \dots  \,,
\ee
where $\lambda$ is the fermion component of the superfield $T$, which is explicitly defined as 
\be
\lambda_\alpha = \frac{1}{\sqrt 2} D_\alpha T | \,.  
\ee
In order to find the flow of $q$ we need the superpotential part that enters $L_{\text{int.}}$, namely, in terms of dimensionless fields,
\be 
L_{\text{int.}} = 
- X \lambda^\alpha G_\alpha 
+ \overline X \overline \lambda^{\dot \alpha} \overline G_{\dot \alpha} + \dots \,.
\ee 
The right hand side of the ERG equation is then 
\be 
\label{FermFermB}
\begin{aligned}
&  -i \int d^4 \hat k (2 \pi)^4 \hat k^{-2} \dot c(\hat k) \, \overline \sigma^{m \dot \alpha \alpha} \hat k_m 
\, \frac{\partial L_{\text{int.}}}{\partial \overline \lambda^{\dot \alpha}(- \hat k)} \frac{\partial L_{\text{int.}}}{\partial \lambda^\alpha(\hat k)} =
\\
= & \ i \int \frac{d^4 \hat p_1 d^4 \hat p_2 d^4 \hat p_3 d^4 \hat p_4}{(2 \pi)^{12}}  
 \int d^4 \hat k \ \frac{\dot c(\hat k)}{\hat k^{2}} \  \hat k^{\dot \alpha \alpha} 
\left( \overline X (-\hat p_1) \overline G_{\dot \alpha} (-\hat p_2) \right) \left( X(\hat p_3) G_\alpha(\hat p_4) \right) \delta_{1,2,\hat k} \delta_{3,4,\hat k} 
 \\
= & \ - i \int \frac{\prod_{i=1}^4 d^4 \hat p_{i}}{(2 \pi)^{12}} 
 \ \frac{\dot c(\hat p_3 + \hat p_4)}{(\hat p_3 + \hat p_4)^{2}} \ 
(\hat p_3 + \hat p_4)^{\dot \alpha \alpha} 
\left( \overline X (-\hat p_1) \overline G_{\dot \alpha} (-\hat p_2) \right) 
\left( X(\hat p_3) G_\alpha(\hat p_4) \right) \delta_{1,2,-3,-4} 
\\
 = & \ - 2 c_1  \int \frac{d^4 \hat p_1 d^4 \hat p_2 d^4 \hat p_3 d^4 \hat p_4}{(2 \pi)^{12}}  \, 
\overline \sigma^{m \dot \alpha \alpha} i (\hat p_3 + \hat p_4)_m \, \left( \overline X (-\hat p_1) \overline G_{\dot \alpha} (-\hat p_2) \right) 
\left( X(\hat p_3) G_\alpha(\hat p_4) \right) \delta_{1,2,-3,-4} \,, 
\end{aligned}
\ee
where we have abbreviated $\hat l^{\dot \alpha \alpha} = \overline \sigma^{m \dot \alpha \alpha} \hat l_m$ in the middle lines. 
The left hand side of \eqref{ERGCC}, always in terms of dimensionless fields and couplings, is then 
\be 
\dot L_{\text{int.}} = (\dot \zeta + 2 \zeta) 
\int \frac{\prod_{i=1}^4 d^4 \hat p_{i}}{(2 \pi)^{12}}  
\overline X(-\hat p_1) \overline  G_{\dot \alpha}(-\hat p_2) \overline \sigma^{m \dot \alpha \alpha } i(\hat p_3+\hat p_4)_m G_\alpha(\hat p_3) X(\hat p_4) 
\delta_{1,2,-3,-4} \,, 
\ee 
so that we can deduce the equation
\be
\dot \zeta = - 2 \zeta - 2 c_1 \,, 
\ee
which is in exact agreement with the result coming from the auxiliary field potential. 
Further cross-checks for the flow can be done. 
It is actually straightforward to show that all the component terms in $\int d^4 \theta |X|^2$ share the same flow, and similarly for $\int d^4 \theta |T|^2$. 
Let us note in passing that it would be interesting to perform a similar analysis for a Volkov--Akulov model in lower dimensions where different flow equations could apply (see e.g. \cite{Feldmann:2017ooy}).

\subsection{What are the bound states?}

Here we wish to determine the nature of the composite states described by $X$ and $T$.
The scalar $X$ is clearly a multi-linear bound state of goldstini, which is controlled by $G^2/2F^X$, which contains a goldstino bilinear,
\be
\label{X-G}
\langle X \rangle \ \sim \ \Big{\langle} \frac{G^2}{f} \Big{\rangle} + \dots \,, 
\ee
taking into account that $F^X$ has an on-shell expansion in terms of the goldstino.\\ 
The scalar $T$ is also multi-linear in the goldstini in the on-shell Volkov--Akulov theory. 
To see this let us first take the superspace equations of motion for $X$, before imposing the nilpotency condition via $T$, which give 
\be
\label{EQX}
- \frac14 \overline D^2 \overline X = - f - TX \,. 
\ee
On this superspace equation one can then impose the condition $X^2=0$, because it is derived from the independent variation of $T$. 
From \eqref{EQX} we see that, once we project to the lowest components, we have $TX = - f + \overline D^2 \overline X | /4 = -f - \overline F^X$ and, using the on-shell value of $F^X$, we find 
\be
\label{TEXP}
T \times G^2 = G^2 
\left(1 - \frac{1}{4 f^4} \overline G^2 \partial^2 G^2  \right) 
\left( \frac{1}{2 f^2} \partial^2 \overline G^2 
+ \frac{3}{8 f^6} \overline G^2 \partial^2 G^2  \partial^2 \overline G^2   \right) \,. 
\ee
This equation shows that the on-shell value of $T$ in the Volkov--Akulov model can be determined in terms of goldstino multi-linears, up to a $G^2$ ambiguity, and it has the form 
\be
\label{T-G}
\langle T \rangle \ \sim \ \Big{\langle} \frac{\partial^2 \overline G^2  }{f^2} \Big{\rangle} + \dots \,.
\ee
Clearly, both \eqref{X-G} and \eqref{T-G} can be recast in other forms due to the on-shell properties of $G$ and the inherent ambiguity of $T$. As it was noticed in \cite{Kuzenko:2011tj}, this ambiguity comes from the fact that the superpotential \eqref{SW} remains invariant under the shift $T \to T + {\cal W} X$, for any holomorphic ${\cal W}$. 
Finally, the fermion component of the superfield $T$, which is $\lambda$, can be found by simply applying a supersymmetry transformation on \eqref{T-G}.

\section{Consequences for the pure Volkov--Akulov model}

\subsection{Critical point stability analysis}

To analyze the properties of the model at a lower energy scale, we refer to \eqref{Knorm} and \eqref{Wnorm} and use the regulator $c(\hat{p}^2)=(1-\hat{p}^2) \Theta(1-\hat{p}^2)$, which gives $c_1 = -1$ and $N = -\frac{1}{32\pi^2}$. We stress once again that we can make this choice without loss of generality as far as the lower energy dynamics are concerned. The K\"ahler potential and the superpotential become 
\be \label{KLT}
\begin{aligned}
K \equiv & |X|^2 + |T|^2 + \tilde{\zeta} |X|^4 + \tilde{\gamma} |X|^2 |T|^2 =\\
= & \ |X|^2 + |T|^2 + \frac14 \, \frac{(1-e^{-2t})}{\left[1 - \frac{1}{16\pi^2} + \frac{1}{8\pi^2} (t + \frac{1}{2} e^{-2t}) \right]^2} |X|^4  +\\ &+ \frac{(1-e^{-2t})}{\left[1 - \frac{1}{16\pi^2} + \frac{1}{8\pi^2} (t + \frac{1}{2} e^{-2t}) \right] \left[-\frac{1}{32\pi^2} + \frac{1}{16\pi^2}(t + \frac12 e^{-2t})\right]}  |X|^2 |T|^2 \, 
\end{aligned}
\ee
and 
\be \label{WLT}
\begin{aligned}
W \equiv& \tilde{f} X + \tilde{g} X^2 T =\\ = &  \ \frac{e^{2t} \xi_{UV}}{\left[1 - \frac{1}{16\pi^2} + \frac{1}{8\pi^2} (t + \frac{1}{2} e^{-2t}) \right]^{1/2}} X  +\\
&+ \frac12 \frac{1}{\left[1 - \frac{1}{16\pi^2} + \frac{1}{8\pi^2} (t + \frac{1}{2} e^{-2t}) \right] \left[-\frac{1}{32\pi^2} + \frac{1}{16\pi^2}(t + \frac12 e^{-2t})\right]^{1/2}} 
X^2 T \,, 
\end{aligned}
\ee
where the couplings $\tilde{\zeta} = \frac{\zeta}{4\alpha^2}$, $\tilde{\gamma} = \frac{\gamma}{\alpha \beta}$, $\tilde{f} = \frac{e^{2t} \xi_{UV}}{\sqrt{\alpha}}$ and $\tilde{g} = \frac{1}{2 \alpha \sqrt{\beta}}$ are obtained by canonically normalizing the fields, i.e. dividing by appropriate powers of the wavefunction renormalization. 

The scalar potential is defined as
\be \label{VLT}
V = g^{i \overline j} \partial_i W \partial_{\overline j} \overline W \,,
\ee
where the indices $i, j, ... = 1, 2$ run over the complex scalar fields $X$ and $T$ and $g^{i \overline j}$ is the inverse of the scalar field space metric $g_{i \overline j} = \partial_i \partial_{\overline j} K$. 

Once $X$ and $T$ (and their complex conjugates) are expressed in terms of real scalar fields via
\be
X = \frac{1}{\sqrt{2}}(\phi + i \chi) \,, \quad T = \frac{1}{\sqrt{2}}(\tau + i \sigma) \,,
\ee
it can be easily shown that the scalar potential has a critical point at
\be \label{centralvacuum}
\phi\Big{|}_{*} = 0 \, , \quad \chi\Big{|}_{*} = 0 \,, \quad \tau\Big{|}_{*} = 0 \, , \quad \sigma\Big{|}_{*} = 0 \,,
\ee
where its value is
\be
V\Big{|}_{*} = \frac{e^{4t} \xi_{UV}^2 }{1 - \frac{1}{16\pi^2} + \frac{1}{8\pi^2} (t + \frac{1}{2} e^{-2t})} \ .
\ee
This is a positive energy configuration, where supersymmetry is spontaneously broken. 

The (in)stability of such critical point is deduced from the scalar mass matrix associated with $V$, evaluated on the configuration itself. The (multiplicity two) eigenvalues are
\begin{equation} \label{ScMass}
\begin{aligned}
    m_{\pm}^2 = - \tilde{f}^2\left[ \left(\tilde{\gamma} + 4 \tilde{\zeta}\right) \pm \sqrt{\frac{16 \ \tilde{g}^2}{\tilde{f}^2} + \left(\tilde{\gamma} - 4 \tilde{\zeta}\right)^2} \ \right] \,.
\end{aligned}    
\end{equation}
As it can be clearly seen, at least one of the eigenvalues \eqref{ScMass} is negative: the scalar field space origin is always tachyonic. We can further observe that, as the RG time flows, the term $\left(\tilde{g}/\tilde{f}\right)^2$ decreases, rendering all the eigenvalues negative for sufficiently large $t$ ($t \gtrsim 0.35$ for $\xi_{UV} = 1$), provided $\tilde{\gamma}$ and $\tilde{\zeta}$ are positive, which is the case. Choosing $\xi_{UV}$ larger will simply increase $\tilde{f}$ and make the tachyonic behaviour more extreme.

The investigation of the existence of other critical points and the possible consequent discussion of their relevance can be simplified by exploiting the following observation. 
The K\"ahler potential and the superpotential have a R-symmetry under which the superfields $X$ and $T$ have opposite non-vanishing R-charges. 
This means that the scalar potential will have a R-symmetry, even though the latter may or may not be preserved by all the other (e.g. higher derivative) interactions. Therefore, once we leave the central configuration, one of the scalar fields is bound to behave like a R-axion Goldstone mode, at least as far as the scalar potential is concerned. 
Then, by definition, such a mode will be massless and it will have a shift symmetry {\it on} the critical points that are away from the central one. Therefore, we can set it to vanish without loss of generality. 
We can consistently choose $\sigma$, which is bound to contribute to the R-axion as long as $\langle \tau \rangle \ne 0$, to be
\be
\sigma = 0 \,.
\ee
With this choice, we see that in order for the gradient of $V$ to be able to vanish, $\chi$ has to be set to zero too. We can thus restrict to the $\phi$ and $\tau$ directions to search for other possible critical points. 
As long as 
\be
t > \frac{1}{2} \log{\frac{1+48\pi^2+\sqrt{1+224\pi^2+2304\pi^4}}{64\pi^2}} \sim 0.20 \quad \text{(for } \xi_{UV}=1\text{)}\,,
\ee
a positive energy critical point, which is also tachyonic, can be found. Moreover, for larger $t$, the characteristic field values for this critical configuration are $\phi^3 \sim \tilde{f}/\tilde{g}$ and $\tau^3 \sim \tilde{f}^2 /\tilde{g}^2$. Since $\tilde{f}$ grows exponentially with $t$, coherently with the small field approximation we are working with, these critical points become untrustable very soon (already for $t \sim 1$) along the RG flow.

\begin{figure}
\includegraphics[scale=0.5]{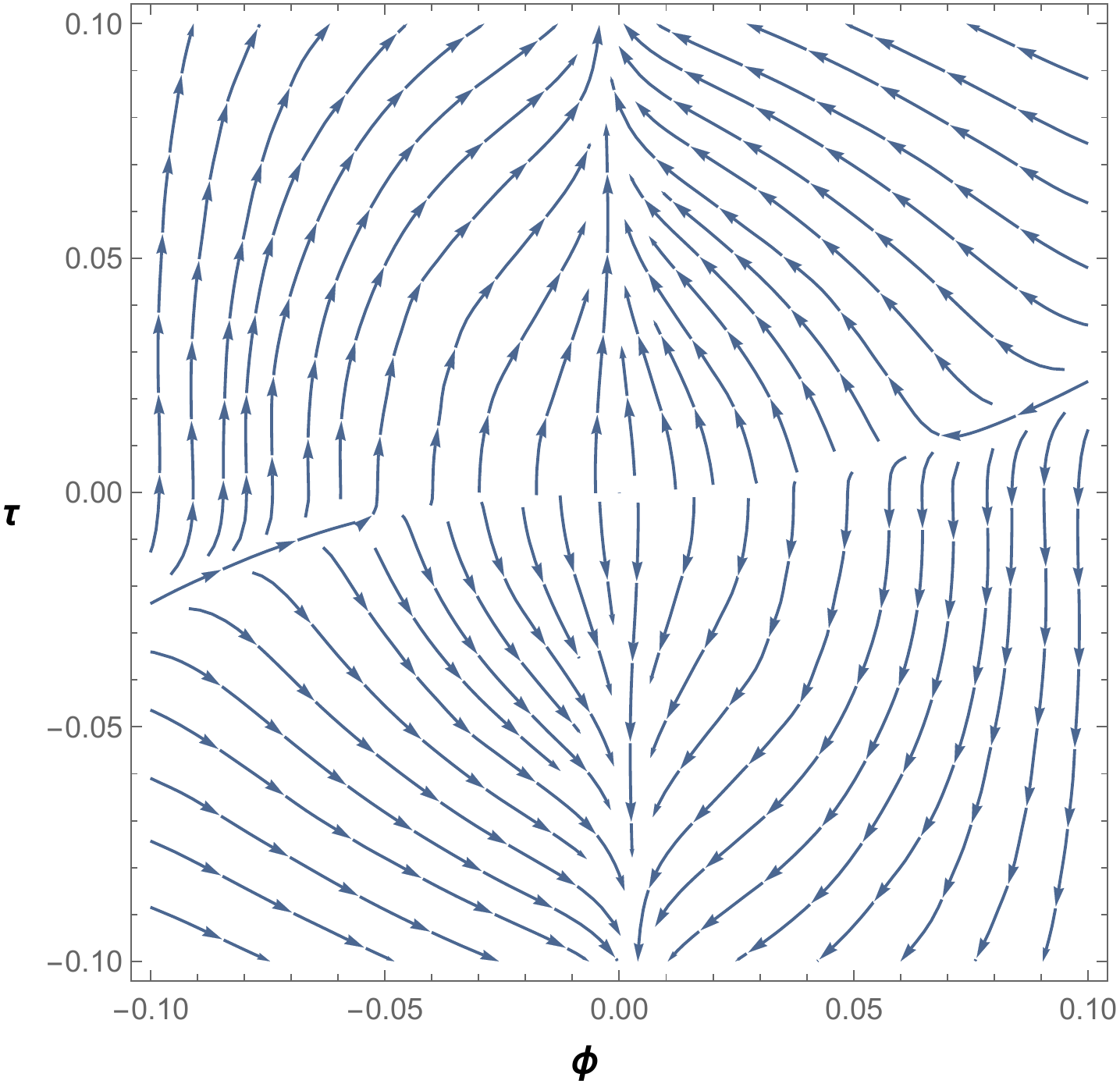} ~~~~~
\includegraphics[scale=0.5]{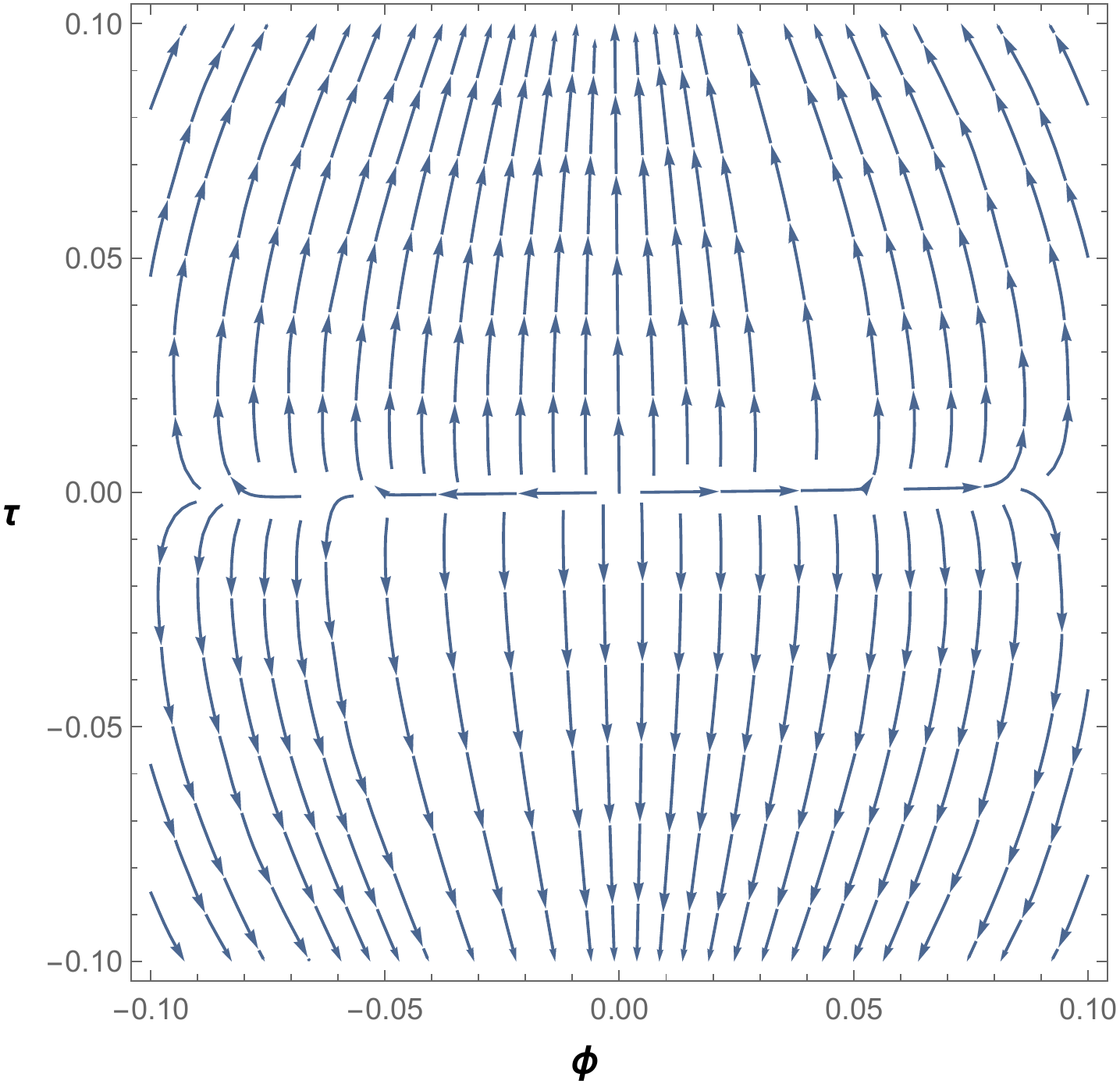}
\centering
\caption{The figure shows the stream-plots of the (negative) gradient of $V$ at $t = 0.1$ (on the left hand side) and $t=1$ (on the right hand side) restricted to the $\phi$ and $\tau$ directions, after consistently setting $\chi$ and $\sigma$ to zero, for $\xi_{UV} = 1$. The origin has positive energy and has one tachyonic direction immediately, while a second tachyonic direction develops as early as $t\gtrsim 0.35$. The behaviour of the masses does not change for larger $t$.} 
\label{SPt80} 
\end{figure}

To give a flavour of the behaviour of the scalar potential $V$ along the non-axionic directions $\phi$ and $\tau$, we include in Figure \ref{SPt80} the stream plots of the (opposite of the) potential gradient at $t = 0.1$ and $t=1$, for $\xi_{UV} = 1$, where one can observe the appearance of the tachyons.

\subsection{Why is there a tachyon in the central critical point?}

We will now argue that the existence of the central tachyon that we have just encountered is unavoidable for a consistent RG flow.

First, let us note that the superpotential at the UV point (identified by an energy scale, say, $\Lambda_0$) is such that
\be
\label{FERMI-MASS}
W_{ij} \Big{|}_{\text{central point}}  = 0 \,,
\ee
which means that the fermions are bound to remain massless on the central point, located at $T=0=X$. 
If we now assume that, for some reason, a K\"ahler potential that gives positive masses to the scalars $X$ and $T$ has been generated during the flow, then, because the effective masses typically increase as RG time passes, there would be a scale $\Lambda_1$ below which we can remove the scalars $T$ and $X$ and work with a new set of constrained superfields satisfying 
\be
\mu \leq \Lambda_1 < \Lambda_0 \,: \quad X^2 = 0 = XT \,. 
\ee
Let us note that, in contrast to the scalar masses, the masses of the fermions are protected by R-symmetry: therefore, they remain zero. 
Indeed, the R-symmetry assignments are 
\be
[G_\alpha]_R = -1 \,, \quad [\lambda_\alpha]_R = 3 \,, 
\ee
and there is no mass combination that can yield a R-invariant. 
The only way in which a scalar could remain massless is if it was a R-axion, but on the central critical point the R-symmetry is unbroken.

Let us repeat the same procedure by decreasing the energy scale $\mu$ below $\Lambda_1$ and find the new effective theory. 
Initially, the superpotential takes the form 
\be
W = f X + \frac12 Y X^2 + \Phi T X \,, 
\ee
and the K\"ahler potential is 
\be
\mu = \Lambda_1 \,: \quad K = |X|^2 + |T|^2 \,, 
\ee
making explicit the role of $Y$ and $\Phi$ as Lagrange multipliers. 
This being fixed, we lower the energy below $\Lambda_1$ and characterize the resulting effective theory. 
We can again observe that the superpotential has the appropriate vertices to generate the kinetic terms for the superfields $Y$ and $\Phi$ and we further notice that, because of the Yukawa couplings, the condition \eqref{FERMI-MASS} is still valid on the central critical point, which is now located at $T=0=X=Y=\Phi$. 
As before, on this configuration the fermions are massless by virtue of the R-symmetry and the scalars are bound to become heavy as we go to lower energies. We can consequently integrate out once more {\it all} the heavy scalars below some energy scale $\Lambda_2$ defining a new effective theory characterized by
\be
\mu \leq \Lambda_2 < \Lambda_1 \,: \quad X^2= 0 = XT = XY = X\Phi  \,.
\ee

The problem is then manifest. 
Unless there is a dynamical reason to stop this procedure, we could get infinite new states in the deep IR, which lead to a possible series of inconsistencies. 
Therefore, the flow self-consistently terminates itself by introducing tachyons that, once appearing, can not be decoupled in a consistent way and, as a consequence, this {\it domino} effect stops.

\subsection{Limitations of the SLPA}

The results that we have obtained above have all been derived in our supersymmetric rendition of the local potential approximation (SLPA), which ignores the generation and feedback of higher derivative terms in the exact renormalization group equations. 
The LPA is a well-motivated and tested approximation and it is a common practice in ERG calculations, while the SLPA is a minimal modification of it, motivated by supersymmetry.
However, it is important to discuss its regime of validity and the possible corrections that one could expect to our results because of its use. 

Since we start from a UV model that has vanishing non-K\"ahler interactions, the higher derivative terms have to be generated before feeding back into the flow for the scalar potential and their effect is expected to appear at higher order in the RG time. The SLPA can be regarded as accurately giving the RG flow for a small decrease in the energy scale. 
A way to see this consists of solving the full exact renormalization group recursively by discretizing $t$ and starting from the UV values of the couplings, that is by making reference to the K\"ahler potential $K=|X|^2$ and the standard superpotential \eqref{SW}. 
In addition, for concreteness, we will consider the flow with the use of the optimized regulator $c(\hat p^2) = \left( 1 - \hat p^2 \right) \Theta\left( 1 - \hat p^2 \right)$ and we will keep the propagator pieces for both $T$ and $X$. 
In the first step one would generate the SLPA terms $\int d^4 \theta |X|^4$ and $\int d^4 \theta |X|^2 |T|^2$ and, on top of them, the higher derivative term $\int d^4 \theta T \partial^2 \overline T$, which is quadratic in the auxiliary fields, together with the higher derivative term $\int d^2 \theta X^2 \partial^2 T$, which is linear in the auxiliary fields. 
Such higher derivative terms are ignored in the SLPA. 
The next recursive step would immediately give the wavefunction renormalization of $X$ and $T$, due to the SLPA effect of $\int d^4 \theta |X|^4$ and $\int d^4 \theta |X|^2 |T|^2$, thus making the composite states manifest. Conversely, the effect of the higher derivative terms would still be inconsequential for the K\"ahler potential. 
Indeed, it takes additional steps in this recursive approach until the higher derivative terms start to backreact on the dominant SLPA contributions including the wavefunction renormalization. 
This is a reflection of the fact that our SLPA approach does not keep track of the anomalous dimension\footnote{A version of the LPA that incorporates the anomalous dimension has been suggested for example in \cite{Bervillier:2014tla}. It would be interesting to see if a similar modification can be made for the SLPA.}, and thus one may not trust the approximation quantitatively for large $t$, where its effect might possibly alter the flow.
If one instead considers an infinitesimal $t$, the effective theory of the composite states with a new cut-off, which is infinitesimally near the one corresponding to the start of the flow (where the SLPA dominates), 
\be
\Lambda_{\text{New}} \lesssim \Lambda_{VA} \,,
\ee
can be derived. 
This means that, in any case, we get the description of the Volkov--Akulov model in terms of a new EFT defined with a slightly lower cut-off. 

Even though we are only slightly moving away from the UV point along the RG flow, there are two features of our SLPA results that we can argue remaining robust even for larger $t$. 
These are the dynamic nature of the superfield $T$ and its tachyonic behaviour near the central critical point.\\ 
The tachyonic nature of the central point has actually two sources. 
The first one is the off-diagonal $X$-$T$ terms of the scalar mass matrix, due to the superpotential, which always give tachyons and dominate at small $t$. On top of that, the second source of instability lies in the positivity of the couplings $\zeta$ and $\gamma$, as it can be seen from \eqref{ScMass}, which dominates at large $t$. 
For both $\zeta$ and $\gamma$ the flow derived in the framework of the SLPA depends only on themselves and results in a monotonically increasing flow. 
Thus, the only way to remove the central tachyons would be for the higher order corrections to overpower the SLPA contribution. 
This would indicate that the theory has reached a point in the RG flow where higher derivative terms are large enough to compete with lower derivative ones. 
As far as the contribution to the instability of the central critical point due to the superpotential (which does not change during the SLPA flow) is concerned, it has to be overcome by some stabilizing contribution from $\zeta$ and $\gamma$. 
This will again lead to the aforementioned intricacies.\\ 
A similar argument can be made regarding the dynamic nature of $T$. Given that at small $t$, where the SLPA can be trusted, the $\gamma$ coupling is positive, $T$ clearly acquires a positive kinetic term. Once this happens, it is impossible for the flow to bring this kinetic term back down for any finite energy, because it would require significant effects from higher derivative terms, breaking the EFT description. Moreover, if the corrections managed to pull the kinetic term of $T$ back to zero at any finite energy, at energies below that we would run the risk of obtaining ghosts.

A possibility that is harder to rule out is that additional couplings appear due to the effects of higher derivative terms that manage to stabilize the potential {\it away} from the central critical point. The SLPA results already indicate the presence of additional (still tachyonic) critical points and additional terms, which could arise from higher derivative contributions, could help in stabilizing them. For instance, a $|T|^4$ term in the K\"ahler potential can be generated via higher derivative terms with its coupling constant parametrically suppressed relative to $\zeta$ and $\gamma$. 
We stress that, even in this case, the conclusion that a goldstino condensate forms remains valid and the dynamics of the theory around this {\it new} vacuum will need to be re-examined.

\section{Coupling to supergravity}

\subsection{Coupling to pure supergravity}

Let us now briefly discuss the supergravity embedding of the model that we have so far presented in the framework of global supersymmetry. 
As a preliminary important comment, we would like to emphasize that the impact of the quantum effects that are related to the supergravity sector is not taken into account here. In other words, we are simply considering the K\"ahler potential and the superpotential (with a possible addition of a constant term) of the composite supersymmetric theory, namely \eqref{Knorm} and \eqref{Wnorm} (or, more specifically, \eqref{KLT} and \eqref{WLT}), coupled to classical supergravity. If the composite fields take values that are parametrically smaller than the cut-off, then the following analysis can be trusted as a first approximation. 

Since we are accessing Supergravity and, therefore, the Planck mass enters as an additional energy scale, we will deal with it in following way. We can set
\be
M_P = \Lambda \times P \,, 
\ee
where a realistic value for the dimensionless parameter $P$ could be, for instance, $P \simeq 10^{4}$. After writing all expressions in terms of dimensionless fields, couplings and momenta, this translates to replacing every instance of $M_P$ by $e^t P$, which is the value of the Planck mass in units of $\mu$. The exponential $e^t$ is actually the ``classical'' flow of $M_P$, simply because it has dimension $[M_P]=1$, as it is the case for any dimensionful coupling that does not flow due to the leading quantum effects that we investigate here. 

As mentioned above, we also slightly modify the superpotential by introducing a constant term, which is related to the Lagrangian gravitino mass
\be
m_{3/2} = e^{\frac{K}{2 M_P^2}} \frac{W}{M_P^2} \,  
\ee
for a K\"ahler potential $K$ and a superpotential $W$.

In accordance with our conventions, we write the superpotential as
\be \label{WLTSUGRA}
\begin{aligned}
W = & \, e^{3t} P^3 W_0 + \frac{e^{2t} \xi_{UV}}{\left[1 - \frac{1}{16\pi^2} + \frac{1}{8\pi^2} (t + \frac{1}{2} e^{-2t}) \right]^{1/2}} X  +\\
&+ \frac12 \frac{1}{\left[1 - \frac{1}{16\pi^2} + \frac{1}{8\pi^2} (t + \frac{1}{2} e^{-2t}) \right] \left[-\frac{1}{32\pi^2} + \frac{1}{16\pi^2}(t + \frac12 e^{-2t})\right]^{1/2}} 
X^2 T \,.
\end{aligned}
\ee
Typically, the superpotential constant term is chosen to be independent of $M_P$, so that the gravity decoupling limit $M_P \to +\infty$ is captured by $m_{3/2}$ approaching $0$; here, however, we measure it directly in Planck units: therefore, $[W_0]=0$. 

We then make use of \eqref{KLT} and \eqref{WLTSUGRA} to calculate the scalar potential
\be \label{VLTSUGRA}
V = e^{e^{-2t} \frac{K}{P^2}} \left(g^{i \overline j} D_i W \, \overline D_{\overline j} \overline{W} - 3 e^{-2t} \frac{W \, \overline{W}}{P^{2}} \right) \,,
\ee
where $D_i W$ is the K\"ahler covariant derivative of $W$,
\be
D_i W = \partial_i W + e^{-2t} \frac{\partial_i K}{P^{2}} W \,.
\ee
If and only if $W_0 = 0$, \eqref{VLTSUGRA} has a de Sitter critical point at
\be
\phi\Big{|}_{*} = \chi\Big{|}_{*} = \tau\Big{|}_{*} = \sigma\Big{|}_{*} = 0 \quad \text{with} \quad V\Big{|}_* = \frac{e^{4t}}{1 - \frac{1}{16\pi^2} + \frac{1}{8\pi^2} (t + \frac{1}{2} e^{-2t})} \quad \text{(for } \xi_{UV} = 1\text{)}\,.
\ee
Computing the scalar mass matrix at such critical configuration, we find that, as in the rigid case, it is {\it highly tachyonic} with masses similar to \eqref{ScMass}, in accordance with the refined de Sitter conjecture  \cite{Obied:2018sgi,Andriot:2018wzk,Garg:2018reu,Ooguri:2018wrx}. 
If $W_0$ is small, the critical point moves away from the origin ($X$,$ T$) $=$ ($0$, $0$) and it still remains highly unstable. Meanwhile, for large enough $W_0$, the potential remains tachyonic, but also gets pulled down to negative energy.

It is also worth noting that our conclusions agree with other results in the literature, and in particular with \cite{Jasinschi:1984cx,Alexandre:2014lla}, where a tachyon shows up in the central critical point. Here, however, we obtain these results in a manifestly supersymmetric setup.

\subsection{Consequences for anti-brane uplifts}

In this section we explore some consequences of the new composite state dynamics and their RG flow for string theory constructions involving anti-brane uplifts of AdS vacua to meta-stable de Sitter critical points. The effect of the anti-brane in these constructions is meant to be captured by adding the Volkov--Akulov K\"ahler potential and superpotential to those of the supergravity model describing the pre-uplift system.
In this way, we assume that the Volkov--Akulov system couples to the other ingredients only via supergravity interactions and the scalar potential for the fields $X$ and $T$, as well as their RG flow, should only be affected by Planck suppressed corrections\footnote{More generally, it is possible that some of the light degrees of freedom that are present in the EFT have more direct coupling to the Volkov--Akulov sector. This includes scenarios which incorporate the effects of warping on the anti-brane studied in \cite{Kachru:2003sx} or the presence of the light complex structure modulus studied in \cite{Bena:2018fqc,Dudas:2019pls}. In this case, the effect of the light fields on the RG flow would, in principle, have to be taken into account. That said, as discussed in section \ref{sec22}, the appearance of the $T$ kinetic term and the tachyonic behaviour that we have described are expected to remain.}. With these assumptions, we can spell out at least two important consequences for models involving nilpotent chiral multiplets.

First, we should expect the tachyonic behaviour near the origin of the ($X$,$T$) field space to remain (and we will verify this explicitly within the KKLT setup). 
The endpoint of this instability will depend on the specifics of additional non-renormalizable operators in the theory. In principle, one expects such corrections to appear from String Theory as well as from corrections to the local potential approximation. This may stabilize the system, but the final configuration is likely to lie at large values of $X$ and $T$ and its physical interpretation is therefore unclear and deserves further investigation.

Second, regardless of the location of the final configuration, one can ask whether it has any chance of remaining at positive energy values. In models with anti-brane uplifts, we note that the final vacuum energy is typically the result of a competition between two dominant terms in the scalar potential
\be \label{vacenergy}
V \sim e^{\frac{K}{M_P^2}} \left( f^2 - M_P^4 V_0 \right) + \dots
\ee
with other contributions being $M_P$-suppressed. Here, $f$ is the coefficient of the linear term in $X$ in the Volkov--Akulov superpotential, while $V_0$ is the pre-uplift contribution to the energy that is independent of the fields $X$ and $T$, but may depend on the other fields in the model. Incorporating the RG flow and working in terms of dimensionless and canonically normalized fields as in the previous sections, the $f^2$ term will flow as
\be \label{Fflow}
f^2 \sim \frac{e^{4t} \xi_{UV}^2}{\alpha(t)} \,,
\ee
where the exponential behaviour corresponds to the ``classical'' RG flow and the $1/\alpha(t)$ factor results from the wavefunction renormalization of the field $X$. The $V_0$ term will, of course, also have the ``classical'' exponential growth; however, it will not inherit the wavefunction renormalization of the fields $X$ or $T$. At this point we note that $\alpha(t)$ is a monotonically increasing function, and thus the uplift term in the potential becomes suppressed at lower energies. If the superpotential contains a constant term, as it is typical in most models of moduli stabilization, its contribution to the potential will not receive any additional suppression. This means that there is a tendency for $V_0$ to dominate over the uplift term at lower energies, possibly resulting in an AdS vacuum. It therefore appears that, in order for the uplift term to remain ``competitive'' at lower energies, the superpotential can not contain terms that are independent of any degrees of freedom in the low energy effective theory. This also means that any heavy moduli that are integrated out must not have VEVs contributing to the superpotential.

\begin{figure}
\includegraphics[scale=0.48]{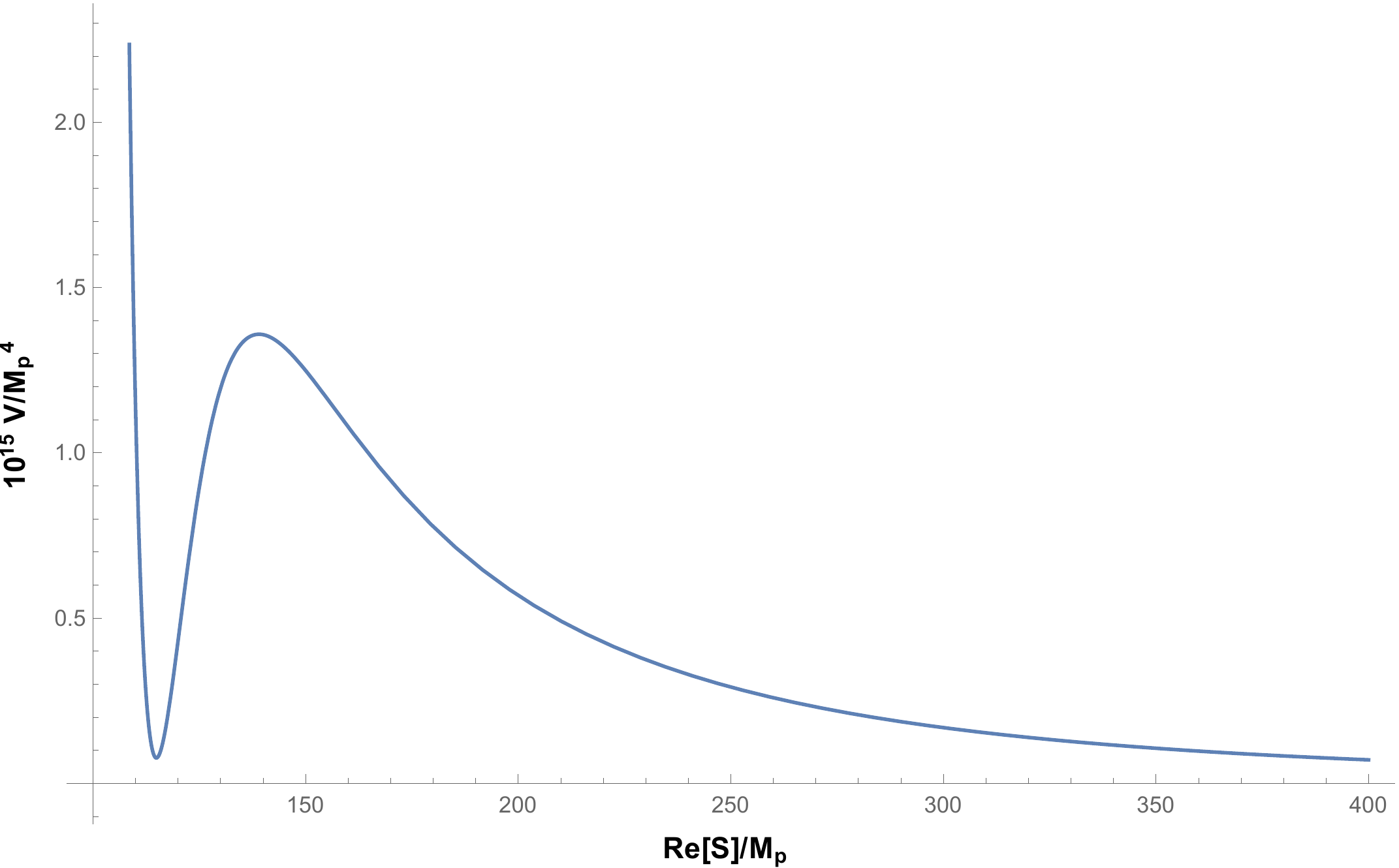}
\centering
\caption{The figure shows the KKLT scalar potential obtained from \eqref{compokklt} for $X=T=0$ and $A=1$, $a=0.1$, $W_0 = -10^{-4}$, $P = 80 \sqrt{\xi_{UV}}$ at the UV cut-off of the effective theory ($t=0$).}
\label{kkltpotential}
\end{figure}

Note that in our approach $\alpha(t)$ grows linearly at large $t$. This behaviour, however, will be corrected by the anomalous dimension of $X$, which our approach ignores. As a first check, we can naively insert a small anomalous dimension $\delta$ into the flow equation for $\alpha$, giving
\be
\dot{\alpha} = - \delta \alpha - 2 N(\gamma + \zeta).
\ee
For negative anomalous dimension, the growth at large $t$ will be more rapid, exacerbating the problem described above, while for positive anomalous dimension the linear growth is expected to stop and approach a finite value, that is of order $(\gamma_{IR}+\zeta_{IR})/\delta$. In this last case, maintaining positive energy might remain possible, but it requires very small $V_0$.

There is another interesting possible \textit{caveat} to the above argument arising from the fact that in a quasi-de Sitter state the Hubble scale provides an IR cut-off, which could potentially halt the RG flow before the negative contributions to the scalar potential overtake the uplift term. Let us imagine a scenario where the scalar potential, evaluated for a particular value of the RG time $t$, has a critical point whose energy is given by an expression of the form \eqref{vacenergy}, with the uplift term flowing as \eqref{Fflow}. The Hubble scale measured in units of the RG scale $\mu$, sourced by this potential, is
\be \label{hubble}
H(t)^2 M_P^2 \sim e^{\frac{K}{M_P^2}} \left(f(t)^2  - M_P^4 V_0 \right) \ \text{ or } \
H(t)^2 P^2 \sim \left(\frac{e^{2t} \xi_{UV}^2}{\alpha(t)} - e^{2t} P^4 V_0 \right) \left(1 + \mathcal{O} \left(\frac{e^{-2t}}{P^2}\right) \right) \nonumber \\
\ee
for large (enough) $t$. Consistency requires that our renormalization scale is above the apparent Hubble scale derived from this potential which can be written as
\be
H(t) = e^{t - t_*}
\ee
with $t_*>t$. The combination of these expressions gives
\be \label{uppertime}
e^{-2 t_*} = \left( \frac{\xi_{UV}^2}{P^2 \alpha(t)} - P^2 V_0 \right) \left( 1 + \mathcal{O}\left(\frac{e^{-2t}}{P^2}\right) \right) \,.
\ee
The condition that $t_* > t$ can potentially put a stop to the RG flow, provided that the above equation can be satisfied when setting $t = t_*$.
For $V_0=0$ there is always a solution as long as $\alpha(t)$ does not grow exponentially, since the exponential on the left hand side of 
\eqref{uppertime} overpowers the sub-exponential growth of $\alpha(t)$ and both sides of the equation asymptote to zero. For small enough $V_0$, a solution continues to exist and pushes $t_*$ higher. In either case, the RG flow will eventually stop at a finite value of $t$, resulting in a de Sitter critical point with a Hubble scale that is exponentially suppressed relative to the UV cut-off.\\
On the other hand, for sufficiently large $V_0$ the large $t$ solution to \eqref{uppertime} disappears entirely, meaning that the IR cut-off disappears, and the effects of the anomalous dimension of $X$ remain the only potential mechanism of staying at positive energy.

\begin{figure}
\includegraphics[scale=0.53]{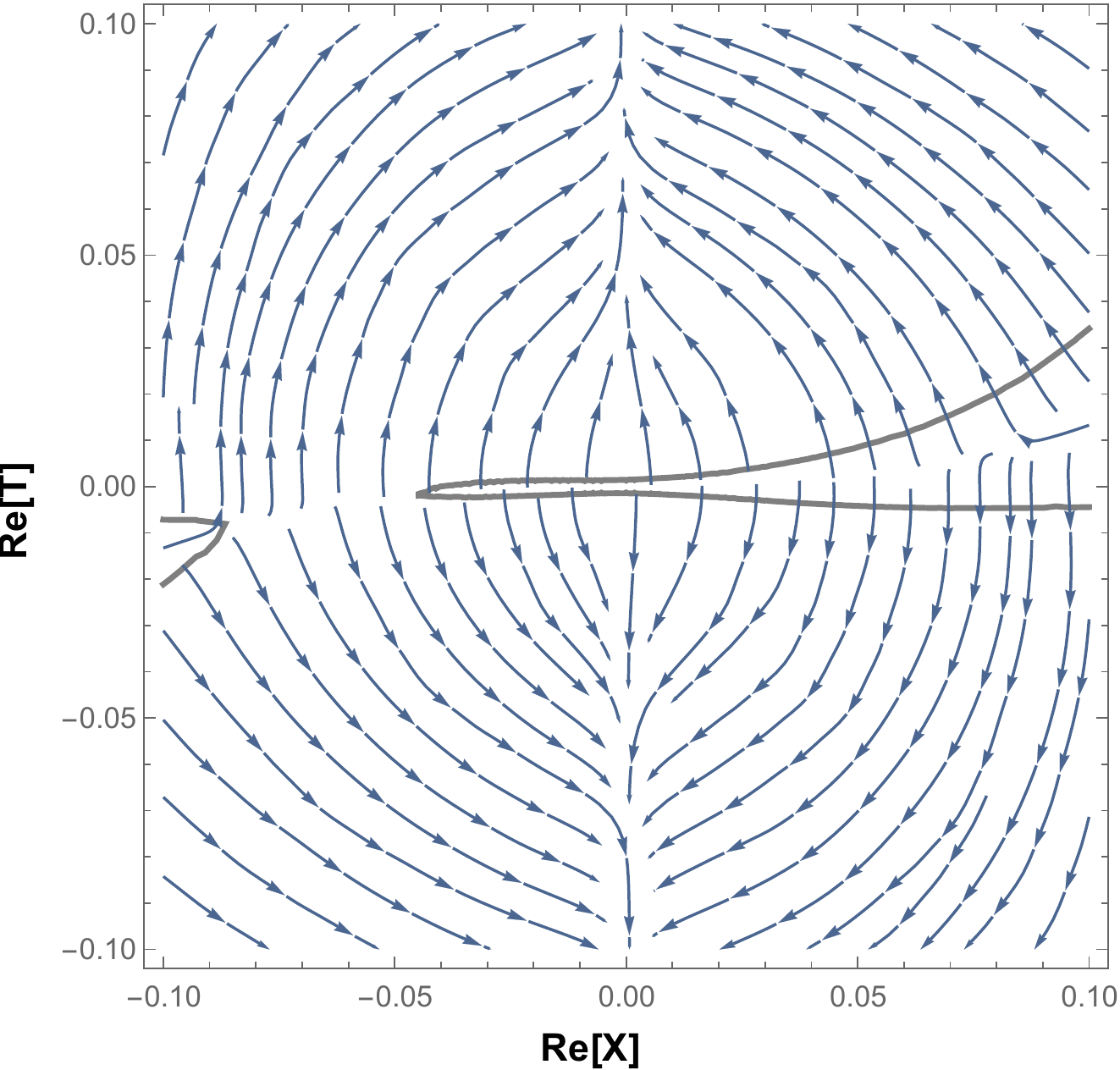}
\centering
\caption{This figure presents the stream-plot of the (negative) gradient of the scalar potential with RG time $t=0.1$ for $S=114.92$ and $A=1$, $a=0.1$, $W_0 = -10^{-4}$, $\xi_{UV}=1$ and $P = 80$ near the origin of the ($X$,$T$) field space. The black contour shows the location where the $\text{Re}S$ component of the gradient vanishes as well. The de Sitter critical point is slightly shifted to ($S=114.9$, $X=-0.046$, $T=-0.0017$) and is tachyonic in the $T$ direction.}
\label{newcrit1}
\end{figure} 

We can investigate these effects in the familiar KKLT setup. We couple our description of the Volkov--Akulov model to an additional chiral multiplet $S$ governed by the pre-uplift KKLT K\"ahler potential and superpotential. This means that we have $K=-3 M_P^2 \log[(S + \bar{S})/M_P] + |X|^2$ and $W= W_0 + A e^{- a S / M_P} + fX$, where $X^2=0$. 
As mentioned above, we will assume that the additional supergravity couplings do not greatly affect the RG flow for the $X$ and $T$ couplings as well as only include the ``classical'' running for the couplings in the $S$ sector.

In our dimensionless conventions the total K\"ahler potential and superpotential now take the form 
\be \label{compokklt}
\begin{aligned}
K = -3 P^2 e^{2t} \log \left(\frac{S + \bar{S}}{P e^t }\right) + K_{norm.} \\
W =  P^3 e^{3t} \left(W_0 + A e^{-\frac{a S}{P e^t}}\right) + W_{norm.}
\end{aligned}
\ee
with $K_{norm.}$ and $W_{norm.}$ given in \eqref{Knorm} and \eqref{Wnorm}, respectively. In this form, the parameters $W_0$, $A$ and $a$ are expressed in Planck units. The strength of the uplift is governed by the ratio of $P/\sqrt{\xi_{UV}}$. For instance, the original example given in \cite{Kachru:2003aw}, where the uplift term was $D/(\text{Re}S)^3$ with $D=3 \times 10^{-9}$ corresponds in our conventions to $P = 80.34 \sqrt{\xi_{UV}}$. The value of $\xi_{UV}$ itself expresses the supersymmetry breaking scale in units of the starting UV cut-off, where we impose that the kinetic term of $T$ vanishes.

\begin{figure}
\includegraphics[scale=0.5]{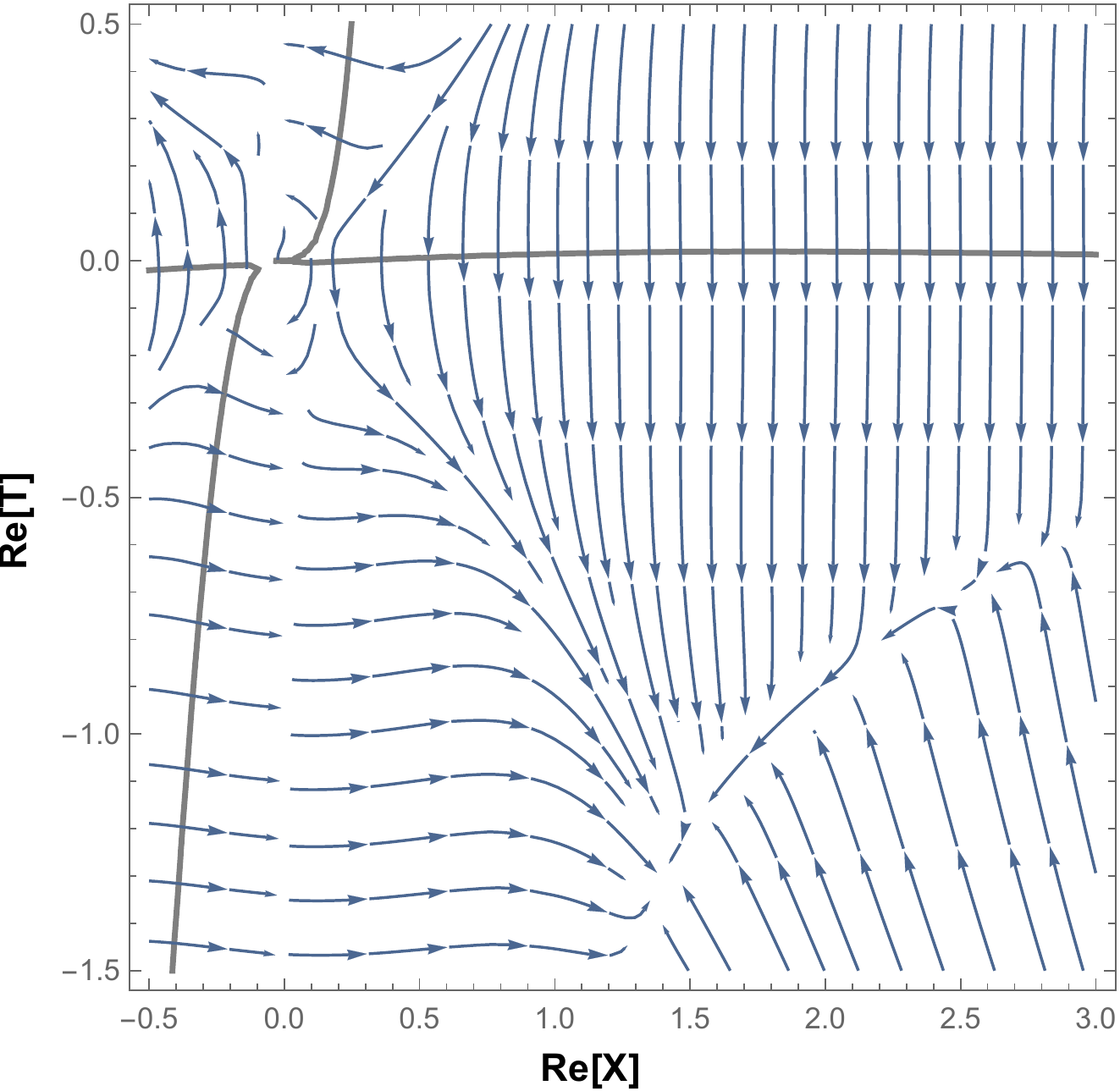}
\centering
\caption{The figure shows the stream-plot of the (negative) gradient of the scalar potential with RG time $t=0.1$ for $S=114.9$ and $A=1$, $a=0.1$, $W_0 = -10^{-4}$, $\xi_{UV}=1$ and $P = 80 $ near the origin of the ($X$,$T$) field space. The black contour shows the location where the $\text{Re}S$ component of the gradient vanishes as well. The ``minimum'' at $X=1.42$, $T=-1.27$ no longer has a non-vanishing gradient along the $S$ direction.}
\label{newcrit2}
\end{figure}

At our UV cut-off, where the field $T$ becomes non-dynamical and imposes the nilpotency condition on $X$, we recover the standard KKLT scenario, with the uplift realized via nilpotent superfields. For suitable choices of parameters, one obtains a potential with the familiar de Sitter meta-stable minimum (see Figure \ref{kkltpotential}). 
However, evolving the theory down the RG flow even slightly, the field $T$ becomes dynamical and the nilpotency condition on $X$ is relaxed. The de Sitter critical point moves slightly in the ($X$,$T$) plane. More importantly, this critical point is not stable in the $X$ and $T$ directions, but develops a tachyon roughly along the $T$ direction (see Figure \ref{newcrit1}). As in the rigid case, the tachyonic behavior appears for all values of $\xi_{UV}$ and only becomes worse as we increase it. 

The endpoint of this tachyonic instability is unclear; however, even by considering the potential for fixed $S$, we can see that it rolls down to a configuration with negative energy at field values of $\mathcal{O}(1)$, where possible higher order terms in the K\"ahler potential become important (see Figure \ref{newcrit2}). 
The relation of the tachyonic instability that we are finding here to the ``goldstino evaporation'' setup of \cite{Farakos:2020wfc} or the KPV scenario \cite{Kachru:2002gs}, which end in supersymmetric points, is yet unknown. 

\begin{figure}
\includegraphics[scale=0.52]{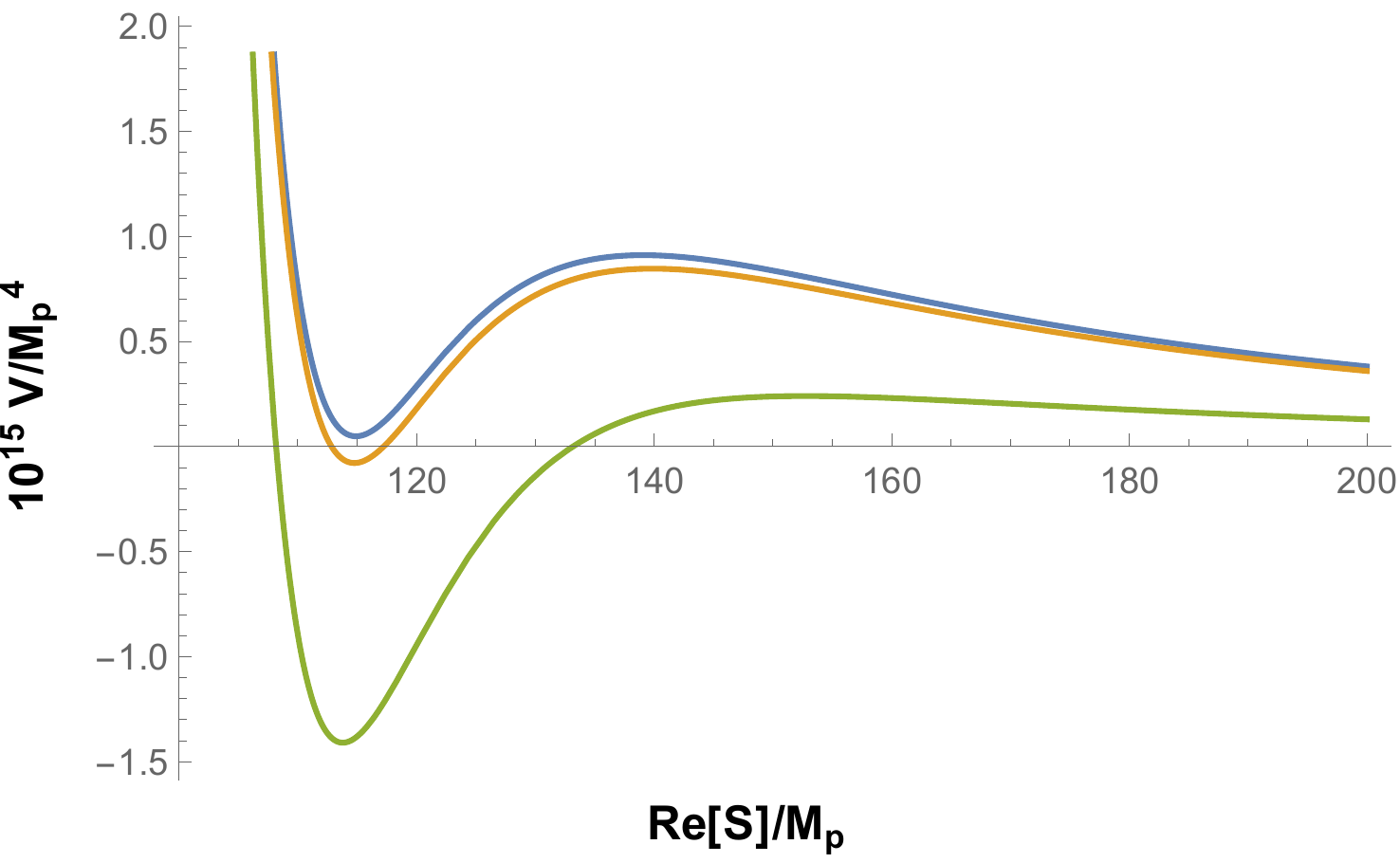}
\centering
\caption{The figure shows the KKLT potential for RG time $t = 0.1\ ,\ 0.3\ ,\ 1$, using the parameter values $A=1$, $a=0.1$, $W_0 = -10^{-4}$, $P = 80$, $\xi_{UV} = 1$ and with ($X$,$T$) restricted to their critical values. The potential at the critical point becomes negative as early as $t\simeq 0.3$.} 
\label{critflow}
\end{figure}

In the above analysis we have tuned the parameters so as to produce the de Sitter critical point in the UV, where the field $X$ is nilpotent. Following the RG flow to a lower energy scale, we find precisely the second problem described above. The linear term in $X$ in the superpotential, which is responsible for the uplift, has the form
\be
W_{\text{up.}} = \frac{e^{2t} \xi_{UV}}{\left[1 - \frac{1}{16\pi^2} + \frac{1}{8\pi^2} (t + \frac{1}{2} e^{-2t}) \right]^{1/2}} X
\ee
and, together with the $W_0$ terms, it represents the main contribution to the scalar potential:
\be
V \sim \frac{1}{|S|^3}\left( \frac{e^{4t} \xi_{UV}^2 }{1 - \frac{1}{16\pi^2} + \frac{1}{8\pi^2} (t + \frac{1}{2} e^{-2t})} - 3 P^4 e^{4t} |W_0|^2\right) + \dots \,.
\ee
As anticipated, we observe the common $e^{4t}$ factor describing the classical running of the potential, and the monotonically increasing denominator due to the wavefunction renormalization of $X$ for the uplift term. The second term receives no such correction by virtue of not containing any dynamical fields that have not been integrated out of the effective theory. Solving \eqref{uppertime} using the typical parameter values $A=1$, $a=0.1$, $W_0 = -10^{-4}$ and $P = 80 \sqrt{\xi_{UV}}$ suggests that a solution might exist for $t \sim 193$, but this is well beyond the small $t$ regime where the SLPA can be trusted. A more careful analysis of the full form of the potential shows that the KKLT critical point reaches negative energies very early in the RG flow, where the SLPA is still valid, as illustrated in Figure \ref{critflow}. Choosing higher values of $\xi_{UV}$ only makes the transition to negative energy happen earlier in the flow. It thus appears that at least for the usual values of the parameters, neither the anomalous dimension of $X$ nor the mechanism for stopping the RG flow via the Hubble scale seem to be able to save the original critical point from reaching negative energies. 

\section{Conclusions}

In this work we analyzed the possible composite states that can be generated from the goldstino in the Volkov--Akulov model due to the fermionic self-interaction. 
Following an exact renormalization group flow we have analyzed the low energy effective theory and we have recast it in a form where supersymmetry becomes linearly realized, albeit spontaneously broken. 
The field space central point that would correspond to the original vacuum turns out to be unstable, and the same happens when we couple the low energy effective theory to 4D $N=1$ Supergravity. 
Similarly, when we couple the system to an additional chiral superfield representing the K\"ahler modulus of KKLT, the instability persists. 
At this stage it is unclear what the origin of this instability in a full string theory setup is, and whether there are alternative models where a stable de Sitter vacuum can exist. 
However, our analysis shows that, when an anti-brane is used for uplift purposes, if the effective description is truly four-dimensional such that 4D $N=1$ non-linear supersymmetry is invoked, stability should not be taken for granted unless the composite goldstino states are first analyzed.

\section*{Acknowledgements}

\noindent
We thank Luca Martucci, Roberto Percacci, Augusto Sagnotti and Luca Vecchi for helpful comments. 
GD is supported in part by MIUR-PRIN contract 2017CC72MK003. 
FF and ME are supported by the STARS grant SUGRA-MAX.

\appendix

\section{ERG equations for chiral supermultiplets}

We derive here the ERG equations for a supersymmetric model involving chiral multiplets. Note that certain conventions and normalizations differ from those of \cite{Polchinski:1983gv}, which we have used in the main text. These will be pointed out when they will occur.

Let us consider the generating functional
\be
\mathcal{Z} (\mu)
= \int \mathcal{D} \Phi \, e^{L[\Phi;\mu]}  
= \int \mathcal{D} \Phi \, e^{L[\Phi;\Lambda]} 
= \mathcal{Z}(\Lambda) \,,
\ee
where $\Lambda$ represents the UV cut-off and $\mu$ is an energy scale of interest below $\Lambda$; $\Phi$ is a collective symbol denoting all the fields of our theory, with individual fields labeled as $\Phi_A$ and $L[\Phi;\mu]$ is the (Euclidean) action at a given energy scale $\mu$.\\
Since n-point correlation functions have to be insensitive to modifications of the energy scale, we require that
\be \label{InvCond}
\dot{\mathcal{Z}}(\mu) \equiv -\mu \partial_{\mu} \mathcal{Z}(\mu) = \int \mathcal{D} \Phi \dot{L}[\Phi;\mu] e^{L[\Phi;\mu]} = 0 \,,
\ee
where $\dot{L}[\Phi; \mu] \equiv -\mu \partial_\mu L[\Phi; \mu]$. In order for \eqref{InvCond} to hold, we require that
\be \label{condition}
\int \mathcal{D} \Phi \dot{L}[\Phi; \mu] e^{L[\Phi; \mu]} = \int \mathcal{D} \Phi \frac{\delta}{\delta \Phi_A}\left( \Psi_A[\Phi; \mu] e^{L[\Phi; \mu]} \right) \,,
\ee
where our convention for the variational derivatives is defined by, say,
\be 
\frac{\delta \Phi_A(p)}{\delta \Phi_B(k)} = (2 \pi)^4 \delta_A^B \delta^{(4)}(p-k) \,
\ee  
(with the normalization of the momentum matching $\delta$-function including a factor of $(2\pi)^4$ relative to the normalization used in the main text), and $\{\Psi_A[\Phi; \mu]\}_A$ are some functionals of the fields. Note that different choices for $\Psi_A$ lead to different parametrizations of the RG flow, but do not change the physics. We then require that
\be \label{ERGbasic}
\begin{aligned}
\dot{L}[\Phi; \mu] = e^{-L[\Phi; \mu]} \frac{\delta}{\delta \Phi_A} \left(\Psi_A[\Phi; \mu] e^{L[\Phi; \mu]} \right) \,,
\end{aligned}
\ee
and try to choose the $\{\Psi_A\}_A$ that result in a convenient expression for the couplings of the various interactions.

\subsection{Bosons}

To proceed, we split the action into a propagator and an interaction piece. Namely, we write the action $L[\Phi;\mu]$ as
\be\label{GendefL}
\begin{aligned}
L[\Phi;\mu] &= L_{\text{prop.}}[\Phi;\mu] + L_{\text{int.}}[\Phi;\mu] = \\ 
&=  \int \frac{d^4 p}{(2\pi)^4}\left[\frac{1}{2} \Phi_A(-p) \left(C^{(\Phi)}_{AB}(p,\mu) \right)^{-1} \Phi_B(p) \right] + L_{\text{int.}}[\Phi;\mu] \,,
\end{aligned}
\ee
where $C^{(\Phi)}_{AB}(p,\mu)$ defines the propagator including a regulator function. We now restrict our attention to $\{\Phi_A\}_A$ being the bosonic fields $\{\phi^a\}_a$ or $\{F^a\}_a$. Since we are assuming massless propagators, we can diagonalize them through appropriate field redefinitions so that
\be
\begin{aligned}
C^{(\Phi)}_{AB}\Big{|}_{\Phi=\phi} &= C^{(\phi)}_{ab} = - \frac{c(p^2/\mu^2)}{p^2}\delta_{ab} \quad \text{for real scalar fields  } \phi^a \,; \\
C^{(\Phi)}_{AB}\Big{|}_{\Phi=F} &= C^{(F)}_{ab} = c(p^2/\mu^2)\delta_{ab} \quad \text{for real auxiliary fields } F^a \,.
\end{aligned}
\ee
Notice that there is a factor of $\sqrt{2}$-difference in the normalization of the auxiliary fields compared to that used in the main text. 
For the real scalar and auxiliary fields this diagonalization allows us to omit the Kronecker-$\delta$ and the species index from the $C^{(\Phi)}$'s. 
The scaling dimensions of these propagators determine the (classical) dimensions of the corresponding fields, which we will denote as $\Delta_{\Phi}$. In momentum space they are $\Delta_{\phi}=-3$ and $\Delta_{F}=-2$. This, in turn, determines the scaling dimensions of the various couplings in $L_{\text{int.}}$. We will denote such couplings as $\{g_{\lambda}\}_{\lambda}$ and their scaling dimensions as $\{\Delta_{\lambda}\}_{\lambda}$.

At this point, let us observe that only the couplings and the propagator have an internal $\mu$ dependence, while the fields and the momenta are $\mu$-independent. We wish, however, to work with dimensionless fields, couplings and momenta and so we carefully track their $\mu$ dependence. The dimensionless quantities are defined as
\be \label{dimLessQuant}
\begin{aligned}
 \hat{p} = p \mu^{-1} \,, \quad  \hat{\Phi}_A(\hat{p}) = \Phi_A(\mu \hat{p}) \mu^{-\Delta_A} \quad \text{and} \quad \hat{g}_\lambda(\mu) = g_\lambda(\mu) \mu^{-\Delta_\lambda} \,.
\end{aligned}
\ee
We also define the dimensionless propagators as
\be
\begin{aligned}
\hat{C}^{(\Phi)}(\hat{p}) = \mu^{-4-2\Delta_{A}} C^{(\Phi)}(p,\mu)
\end{aligned}
\ee
in such a way that the bosonic action propagator term can also be written as
\be
L_{\text{prop.}}[\hat{\Phi}; \mu] = \int \frac{d^4 \hat{p}}{(2\pi)^4}\left[ \frac12 \hat{\Phi}_A(-\hat{p}) \left( \hat{C}^{(\Phi)}(\hat{p}) \right)^{-1} \hat{\Phi}_A(\hat{p}) \right] \,,
\ee
where
\be \label{DimLessProps}
\begin{aligned}
\hat{C}^{(\Phi)}\Big{|}_{\Phi=\phi} &= \hat{C}^{(\phi)} = - \frac{c(\hat{p}^2)}{\hat{p}^2} \quad \text{for real scalar fields  } \phi^a \,; 
\\
\hat{C}^{(\Phi)}\Big{|}_{\Phi=F} &= \hat{C}^{(F)} = c(\hat{p}^2) \quad \text{for real auxiliary fields } F^a \,.
\end{aligned}
\ee
To compute the left hand side of \eqref{ERGbasic}, we will need the $\mu$-derivatives of all the dimensionless quantities that we have introduced. From \eqref{dimLessQuant} we obtain
\be \label{phatdot}
\begin{aligned}
 -\mu \partial_\mu \hat{p} &= - p \mu \partial_\mu \mu^{-1} = p \mu^{-1} = \hat{p}
\end{aligned}
\ee
and similarly for the momentum space measure, $-\mu \partial_\mu d\hat{p} = d\hat{p}$. We also have
\be \label{dimLessDot}
\begin{aligned}
-\mu \partial_\mu  \hat{\Phi}_A(\hat{p}) 
&= -\mu \partial_\mu  \left(\Phi_A(\mu \hat{p}) \mu^{-\Delta_A}\right)
= \Delta_A \mu^{-\Delta_A} \Phi_A(\mu \hat{p}) 
-\mu \Phi_A^\prime(\mu \hat{p}) \partial_\mu (\mu \hat{p}) \mu^{-\Delta_A} 
\\ &= \Delta_A \mu^{-\Delta_A} \Phi_A(\mu \hat{p}) -\mu \Phi_A^\prime(\mu \hat{p}) ( \hat{p} + \mu \partial_\mu \hat{p})  \mu^{-\Delta_A} 
\\ &= \Delta_A \mu^{-\Delta_A} \Phi_A(\mu \hat{p}) -\mu \Phi_A^\prime(\mu \hat{p}) (\hat{p} - \hat{p})  \mu^{-\Delta_A} 
\\
&= \Delta_A \hat{\Phi}_A(\hat{p}) \,;\\
-\mu \partial_\mu c(\hat{p}) &= (-\mu \partial_\mu \hat{p}) \partial_{\hat{p}} c(\hat{p}) = \hat{p} \partial_{\hat{p}} c(\hat{p}) \,,
\end{aligned}
\ee
where $\Phi_A^\prime$ denotes a regular derivative of $\Phi_A$ with respect to its argument.

Armed with these expressions we can compute
\be
\begin{aligned}
\dot{\hat{C}}^{(\phi)}(\hat{p}) &= -(\mu \partial_\mu \hat{p})\partial_{\hat{p}} \hat{C}^{(\phi)}(\hat{p}) = \hat{p} \partial_{\hat{p}}\bigg(-\frac{c(\hat{p})}{\hat{p}^{2}}\bigg) = - \frac{\big(\hat{p}\partial_{\hat{p}} c(\hat{p})\big)}{\hat{p}^2} + 2 \frac{c(\hat{p})}{\hat{p}^2} \,; \\
\dot{\hat{C}}^{(F)}(\hat{p}) &= -(\mu \partial_\mu \hat{p})\partial_{\hat{p}} \hat{C}^{(F)}(\hat{p}) = \hat{p} \partial_{\hat{p}} c(\hat{p}) \,.
\end{aligned}
\ee
Defining, then, 
\be \label{Ctilde}
\tilde{C}^{(\phi)}(\hat{p}) = -\frac{\big(\hat{p}\partial_{\hat{p}} c(\hat{p})\big)}{\hat{p}^2} \ , \quad  \tilde{C}^{(F)}(\hat{p}) = \hat{p} \partial_{\hat{p}} c(\hat{p}) \,,
\ee
we are allowed to write
\be 
\dot{\hat{C}}^{(\Phi)}(\hat{p}) = \tilde{C}^{(\Phi)}(\hat{p}) + (4+2\Delta_{A}) \hat{C}^{(\Phi)}(\hat{p}) \,.
\ee 
From this we can compute 
\be \label{LpropDot}
\begin{aligned}
\dot{L}_{\text{prop.}}^{(\Phi_A)} &= -\mu \partial_\mu \int \frac{d^4 \hat{p}}{(2\pi)^4} \left[ \frac12 \hat{\Phi}_A(-\hat{p}) \left(\hat{C}^{(\Phi)}(\hat{p})\right)^{-1} \hat{\Phi}_A(\hat{p}) \right] = \\
&=\frac12 \int \frac{d^4 \hat{p}}{(2\pi)^4}\left[ \big(4+2\Delta_A\big) \hat{\Phi}_A(-\hat{p}) \left(\hat{C}^{(\Phi)}(\hat{p})\right)^{-1} \hat{\Phi}_A(\hat{p}) -  \hat{\Phi}_A(-\hat{p}) \frac{\dot{\hat{C}}^{(\Phi)}(\hat{p})}{(\hat{C}^{(\Phi)})^2} \hat{\Phi}_A(\hat{p}) \right]\\
&=-\frac12 \int \frac{d^4 \hat{p}}{(2\pi)^4} \left[ \hat{\Phi}_A(-\hat{p}) \frac{\tilde{C}^{(\Phi)}(\hat{p})}{(\hat{C}^{(\Phi)}(\hat{p}))^2} \hat{\Phi}_A(\hat{p}) \right] \,,
\end{aligned}
\ee
so that the left hand side of \eqref{ERGbasic} is
\be
\dot{L} = -\int \frac{d^4 \hat{p}}{(2\pi)^4} \left[\frac12  \hat{\Phi}_A(-\hat{p}) \frac{\tilde{C}^{(\Phi)}(\hat{p})}{(\hat{C}^{(\Phi)}(\hat{p}))^2} \hat{\Phi}_A(\hat{p}) \right] + \dot{L}^{(\chi)}_{\text{prop.}} + \dot{L}_{\text{int.}} \,,
\ee
where in the first term a sum over all scalar and auxiliary fields is understood. 

We now show that for a suitable choice of $\{\Psi_A\}_{A=\phi,F}$ this first term also appears on the right hand side of \eqref{ERGbasic} and do not take part to the final ERG equation. A similar procedure, which we will describe further, allows to get rid of the fermionic propagator terms.

The appropriate choice of $\Psi_A[\hat{\Phi};\mu]$ is as follows:
\be
\Psi_A[\hat{\Phi}; \mu] = \int \frac{d^4 \hat{k}}{(2\pi)^4} \frac12 \tilde{C}^{(\Phi)}(\hat{k}) \frac{\delta \tilde{L}[\hat{\Phi}; \mu]}{\delta \hat{\Phi}_A(\hat{k})}  \quad \text{with} \quad \tilde{L} = - L_{\text{prop.}} + L_{\text{int.}} \,,
\ee
and $\tilde{C}^{(\Phi)}$ being defined by \eqref{Ctilde}. Omitting the functional and function variables of the action, the right-hand side of \eqref{ERGbasic} consequently becomes
\be \label{ERGright}
\begin{aligned}
&e^{-L} \frac{\delta}{\delta \hat{\Phi}_A} \left(\Psi_A e^{L} \right) = \int \frac{d^4 \hat{k}}{(2\pi)^4} \frac12 \tilde{C}^{(\Phi)}(\hat{k}) \left(\frac{\delta^2 \tilde{L}}{\delta \hat{\Phi}_A(-\hat{k}) \delta \hat{\Phi}_A(\hat{k})} + \frac{\delta \tilde{L}}{\delta \hat{\Phi}_A(\hat{k})} \frac{\delta L}{\delta \hat{\Phi}_A(-\hat{k})} \right) =\\
&= \int \frac{d^4 \hat{k}}{(2\pi)^4} \frac12 \tilde{C}^{(\Phi)}(\hat{k}) \left(\frac{\delta^2 L_{\text{int.}}}{\delta \hat{\Phi}_A(-\hat{k}) \delta \hat{\Phi}_A(\hat{k})} + \frac{\delta L_{\text{int.}}}{\delta \hat{\Phi}_A(\hat{k})} \frac{\delta L_{\text{int.}}}{\delta \hat{\Phi}_A(-\hat{k})} +\right. \\
&\left. \qquad \qquad \qquad \qquad \qquad -\frac{\delta^2 L_{\text{prop.}}}{\delta \hat{\Phi}_A(-\hat{k}) \delta \hat{\Phi}_A(\hat{k})} - \frac{\delta L_{\text{prop.}}}{\delta \hat{\Phi}_A(\hat{k})} \frac{\delta L_{\text{prop.}}}{\delta \hat{\Phi}_A(-\hat{k})} \right) \,,
\end{aligned}
\ee 
where no sum over $A$ is implied. The evaluation of the terms in the last line gives
\be 
\begin{aligned}
-\int \frac{d^4 \hat{k}}{(2\pi)^4} \frac 12 \tilde{C}^{(\Phi)}(\hat{k}) \frac{\delta^2 L_{\text{prop.}}}{\delta \hat{\Phi}_A(-\hat{k}) \delta \hat{\Phi}_A(\hat{k})} &= -\int \frac{d^4 \hat{k}}{(2\pi)^4} \frac12 \tilde{C}^{(\Phi)}(\hat{k}) \left(\hat{C}^{(\Phi)}(\hat{k})\right)^{-1}  (2\pi)^4\delta^{(4)}(0)
=\\&= - \frac{1}{2} \int \frac{d^4 \hat{k}}{(2\pi)^4} \frac{\hat{k} \partial_{\hat{k}} c(\hat{k})}{c(\hat{k})} (2\pi)^4\delta^{(4)}(0) \,,
\end{aligned}
\ee
which can be absorbed into the measure, as well as
\be 
-\int \frac{d^4 \hat{k}}{(2\pi)^4} \frac12 \tilde{C}^{(\Phi)}(\hat{k}) \frac{\delta L_{\text{prop.}}}{\delta \hat{\Phi}_A(\hat{k})} \frac{\delta L_{\text{prop.}}}{\delta \hat{\Phi}_A(-\hat{k})} = -\int \frac{d^4 \hat{k}}{(2\pi)^4} \left[\frac12  \hat{\Phi}_A(-\hat{k}) \frac{\tilde{C}^{(\Phi)}(\hat{k})}{(\hat{C}^{(\Phi)}(\hat{k}))^2} \hat{\Phi}_A(\hat{k}) \right] \,,
\ee
which is precisely the term that we need to cancel its analogous on the other side. Putting things together, we obtain the ERG equation
\be \label{ERGfull}
\begin{aligned}
\dot{L}_{\text{int.}}[\hat{\Phi};\mu] = \sum_A \int \frac{d^4 \hat{k}}{(2\pi)^4} \frac12 \tilde{C}^{(\Phi)}(\hat{k}) \left(\frac{\delta^2 L_{\text{int.}}[\hat{\Phi};\mu]}{\delta \hat{\Phi}_A(-\hat{k}) \delta \hat{\Phi}_A(\hat{k})} + \frac{\delta L_{\text{int.}}[\hat{\Phi};\mu]}{\delta \hat{\Phi}_A(\hat{k})}\frac{\delta L_{\text{int.}}[\hat{\Phi};\mu]}{\delta \hat{\Phi}_A(-\hat{k})} \right) \ +\\ +\ \text{fermionic contributions} \,.
\end{aligned}
\ee 

We can now expand $\dot{L}_{\text{int.}}$ so that we can isolate the behaviour of individual coupling constants. As already mentioned, we will denote these couplings as $\{g_{\lambda}\}_{\lambda}$ and their classical scaling dimensions as $\{\Delta_{\lambda}\}_{\lambda}$, where $\lambda$ is an index labelling the various interaction terms. A non-derivative interaction term will have the form
\be
L_{\text{int.},\lambda} = \int \left( \prod_A \frac{d^4 p_A}{(2\pi)^4}\right) g_{\lambda} \prod_A \Phi_A(p_A) \times (2\pi)^4\delta^{(4)}\left(\sum_A p_A \right) \,,
\ee
where $\Delta_\lambda - 4 + \sum_A \left(4+\Delta_{A} \right)  = 0$. (Note that the momentum $\delta$-function has dimension $-4$, in order to properly integrate to unit over momentum space). The conversion to dimensionless quantities gives
\be
\begin{aligned}
L_{\text{int.}, \lambda} &= \int \left( \prod_A \frac{\mu^4 d^4 \hat{p}_A}{(2\pi)^4}\right) \mu^{\Delta_\lambda} \hat{g}_{\lambda} \prod_A \mu^{\Delta_{A}} \hat{\Phi}_A(\hat{p}_A) \times \frac{(2\pi)^4}{\mu^4}\delta^{(4)}\left(\sum_A \hat{p}_A \right) =\\
&= \int \left( \prod_A \frac{d^4 \hat{p}_A}{(2\pi)^4}\right)  \hat{g}_{\lambda} \prod_A \hat{\Phi}_A(\hat{p}_A) \times (2\pi)^4 \delta^{(4)} \left(\sum_A \hat{p}_A \right)  \,,
\end{aligned}
\ee
which is now manifestly dimensionless and all the quantities have the $\mu$ dependence derived above. 
Denoting $-\mu \partial_\mu Q$ as $\dot{Q}$ for some quantity $Q$ and using \eqref{phatdot} and \eqref{dimLessDot} we obtain
\be \label{LdotFinal}
\begin{aligned}
\dot{L}_{\text{int.}, \lambda} &= \int  \left( \prod_A \frac{d^4 \hat{p}_A}{(2\pi)^4} \hat{\Phi}_A(\hat{p}_A) \right) \left[ \left(\sum_A \left(4+\Delta_A \right) - 4\right) \hat{g}_{\lambda} + \dot{\hat{g}}_\lambda \right] \times (2\pi)^4\delta^{(4)} \left(\sum_A \hat{p}_A \right) =\\
&= \int  \left( \prod_A \frac{d^4 \hat{p}_A}{(2\pi)^4} \hat{\Phi}_A(\hat{p}_A) \right) \left( -\Delta_\lambda \hat{g}_{\lambda} + \dot{\hat{g}}_\lambda \right) \times (2\pi)^4 \delta^{(4)}\left(\sum_A \hat{p}_A \right) \,.
\end{aligned}
\ee
The above expression also holds for couplings involving fermions and it is straightforward to show that a similar final expression also holds for higher derivative interactions, but we do not use these in the present work. Overall we have
\be \label{LdotSum}
\dot{L}_{\text{int.}} = \sum_{\lambda}\dot{L}_{\text{int.}, \lambda} \,
\ee
and we can isolate the RG flow of individual couplings by matching terms involving the same field combinations on each side of \eqref{ERGfull}.

\subsection{Fermions}

For a single Weyl fermion we can write the propagator part of the action as
\be
L^{(\chi)}_{\text{prop.}} 
= \int \frac{d^4 \hat{k}}{(2\pi)^4} \chi^{\alpha}(\hat{k}) \hat{C}^{-1}_{ \alpha\dot{\alpha}}(\hat{k})  
\overline{\chi}^{\dot{\alpha}}(-\hat{k})  \,,
\ee
where the fields are dimensionless (and are related to their dimensionful partners by $\Delta_{\chi}=-5/2$), but we omit the hats on them to avoid notation clutter, and we also have 
\be 
\hat C^{-1}_{\alpha \dot \alpha}(\hat k) = -i c^{-1}(\hat k) \sigma_{\alpha \dot \alpha}^m \hat k_m \,.  
\ee 
Note that the Euclidean sigma-matrices satisfy $(\sigma^m \overline \sigma^n + \sigma^n \overline \sigma^m )_\alpha^\beta = 2 \delta^{mn} \delta_\alpha^\beta$.  
The propagator is then 
\be
\hat{C}^{ \dot{\alpha}\alpha}(\hat{k}) = c(\hat{k}) \frac{i \hat{\slashed{k}}^{ \dot{\alpha}\alpha} }{\hat{k}^2} 
\ee
and, as a consequence,
\be
\dot{\hat{C}}^{ \dot{\alpha}\alpha} 
= i \big(\hat{k} \partial_{\hat{k}} c(\hat{k}) \big) \frac{\hat{\slashed{k}}^{ \dot{\alpha}\alpha}}{\hat{k}^2} 
- i c(\hat{k}) \frac{\hat{\slashed{k}}^{ \dot{\alpha}\alpha}}{\hat{k}^2} \,.
\ee
Therefore,
\be
\begin{aligned}
\dot{L}^{(\chi)}_{\text{prop.}} =&  \int \frac{d^4 \hat{k}}{(2\pi)^4} 
\bigg[ \left( 4+2\cdot \left(-\frac52 \right) + 1 \right)\chi^{\alpha}(\hat{k})  \hat{C}^{-1}_{ \alpha \dot{\alpha}}(\hat{k}) \overline{\chi}^{\dot{\alpha}}(-\hat{k}) +  \\
& - \chi^{\alpha}(\hat{k})  \hat{C}^{-1}_{\alpha \dot{\beta}}(\hat{k})
\tilde{C}^{\dot{\beta} \beta}(\hat k) 
\hat{C}^{-1}_{\beta \dot{\alpha}}(\hat{k})\overline{\chi}^{\dot{\alpha}}(-\hat{k}) \bigg] \,,
\end{aligned}
\ee
where
\be
\tilde{C}^{ \dot{\alpha} \alpha} = \bigg[ \big(\hat{k} \partial_{\hat{k}} c(\hat{k}) \big) 
\frac{i \hat{\slashed{k}}}{\hat{k}^2} \bigg]^{ \dot{\alpha} \alpha} \,.
\ee
As in the bosonic case, this term will need to be cancelled by terms coming from
\be
\begin{aligned}
&\int \frac{d^4 \hat{p}}{(2\pi)^4} \left[  \Psi_1^\alpha \frac{\delta L_{\text{prop.}}}{\delta \chi^{\alpha}(\hat{p})}+ \frac{\delta L_{\text{prop.}}}{\delta \overline{\chi}^{\dot{\alpha}}(-\hat{p})} \Psi_2^{\dot{\alpha}}  \right]  = 
\\
&=\int \frac{d^4 \hat{p}}{(2\pi)^4} \left[\Psi_1^{\alpha}  
\bigg(
\hat{C}^{-1}_{ \alpha \dot{\alpha}}(\hat{p}) \overline{\chi}^{\dot{\alpha}}(-\hat{p}) 
\bigg) 
-  \bigg(
\chi^{\alpha}(\hat{p}) \hat{C}^{-1}_{ \alpha \dot{\alpha}}(\hat{p})  
\bigg) \Psi_2^{\dot{\alpha}} \right].
\end{aligned}
\ee
An appropriate choice of $\Psi_{1,2}$ appears to be
\be
\Psi_1^{\beta}(\hat{k}) &= - \frac{1}{2} \chi^{\alpha}(\hat{k}) \hat{C}^{-1}_{ \alpha \dot{\beta}}(\hat{k}) \tilde{C}^{ \dot{\beta}\beta}  + \Psi_{1,\text{int.}} \,; \\
\Psi_2^{\dot{\beta}}(\hat{k})&=\frac{1}{2} \tilde{C}^{ \dot{\beta} \beta} \hat{C}^{-1}_{ \beta \dot{\alpha}}(\hat{k}) \overline{\chi}^{\dot{\alpha}}(-\hat{k}) + \Psi_{2,\text{int.}}
\ee
and by analogy with the bosonic case we consequently arrive at a complete ansatz for $\Psi_{1,2}$:
\be
\begin{aligned}
\Psi_1^{\beta}(\hat{k}) &= - \frac{1}{2}  \frac{\delta \tilde{L}}{\delta \overline{\chi}^{\dot{\beta}}(-\hat{k}) } \tilde{C}^{\dot{\beta} \beta }(\hat{k}) \,; \\
\Psi_2^{\dot{\beta}}(\hat{k})&= - \frac{1}{2} \tilde{C}^{\dot{\beta} \beta }(\hat{k}) \frac{\delta \tilde{L}}{\delta \chi^{\beta}(\hat{k})} \,,
\end{aligned}
\ee
where as before $\tilde{L} = -L_{\text{prop.}} + L_{\text{int.}}$. With these expressions, we have
\be
\begin{aligned}
e^{-L} \frac{\delta}{\delta \chi^\alpha} \big( \Psi_1^\alpha e^{L} \big) 
& = - \frac12 \tilde{C}^{\dot{\alpha} \alpha } \bigg(  \frac{\delta^2 \tilde{L}}{\delta \chi^{\alpha} \delta \overline{\chi}^{\dot{\alpha}}} - \frac{\delta \tilde{L}}{\delta \overline{\chi}^{\dot{\alpha}}} \frac{\delta L}{\delta \chi^{\alpha}} \bigg) =
\\
& = \frac12 \tilde{C}^{\dot{\alpha} \alpha } \bigg(  \frac{\delta^2 \tilde{L}}{ \delta \overline{\chi}^{\dot{\alpha}}  \delta \chi^{\alpha}} + \frac{\delta \tilde{L}}{\delta \overline{\chi}^{\dot{\alpha}}} \frac{\delta L}{\delta \chi^{\alpha}} \bigg) =
\\
& =  \frac12 \tilde{C}^{ \dot{\alpha} \alpha} \left(  \frac{\delta^2 (-L_{\text{prop.}} + L_{\text{int.}})}{ \delta \overline{\chi}^{\dot{\alpha}}  \delta \chi^{\alpha}} + \frac{\delta (-L_{\text{prop.}} + L_{\text{int.}})}{\delta \overline{\chi}^{\dot{\alpha}}} \frac{\delta (L_{\text{prop.}} + L_{\text{int.}})}{\delta \chi^{\alpha}} \right)  
\\
&=  \frac12 \tilde{C}^{\dot{\alpha} \alpha} \bigg( - \hat{C}^{-1}_{ \alpha \dot{\alpha}} (2\pi)^4\delta^{(4)}(0) + \frac{\delta^2 L_{\text{int.}}}{ \delta \overline{\chi}^{\dot{\alpha}}  \delta \chi^{\alpha}} - \frac{\delta L_{\text{prop.}} }{\delta \overline{\chi}^{\dot{\alpha}}} \frac{\delta L_{\text{prop.}}}{\delta \chi^{\alpha}} + \frac{\delta L_{\text{int.}}}{\delta \overline{\chi}^{\dot{\alpha}}} \frac{\delta L_{\text{int.}}}{\delta \chi^{\alpha}} \ + 
\\
& \qquad \qquad - \frac{\delta L_{\text{prop.}} }{\delta \overline{\chi}^{\dot{\alpha}}} \frac{\delta L_{\text{int.}}}{\delta \chi^{\alpha}} +  \frac{\delta L_{\text{int.}}}{\delta \overline{\chi}^{\dot{\alpha}}} \frac{\delta L_{\text{prop.}} }{\delta \chi^{\alpha}} \bigg) 
\end{aligned}
\ee
and
\be
\begin{aligned}
- e^{-L} \frac{\delta}{\delta \overline{\chi}^{\dot{\alpha}} } \big( \Psi_2^{\dot{\alpha}} e^{L} \big) & 
= \frac12 \tilde{C}^{ \dot{\alpha} \alpha} \bigg(  \frac{\delta^2 \tilde{L}}{\delta \overline{\chi}^{\dot{\alpha}} \delta \chi^{\alpha}  } - \frac{\delta \tilde{L}}{\delta \chi^{\alpha} } \frac{\delta L}{ \delta \overline{\chi}^{\dot{\alpha}} } \bigg) =
\\
& = \frac12 \tilde{C}^{ \dot{\alpha} \alpha} \bigg(  \frac{\delta^2 \tilde{L}}{ \delta \overline{\chi}^{\dot{\alpha}}  \delta \chi^{\alpha}} + \frac{\delta L}{\delta \overline{\chi}^{\dot{\alpha}}} \frac{\delta \tilde{L}}{\delta \chi^{\alpha}} \bigg) =
\\
& = \frac12\tilde{C}^{ \dot{\alpha} \alpha} \left(  \frac{\delta^2 (-L_{\text{prop.}} + L_{\text{int.}})}{ \delta \overline{\chi}^{\dot{\alpha}}  \delta \chi^{\alpha}} + \frac{\delta (L_{\text{prop.}} + L_{\text{int.}})}{\delta \overline{\chi}^{\dot{\alpha}}} \frac{\delta ( - L_{\text{prop.}} + L_{\text{int.}})}{\delta \chi^{\alpha}} \right)  \\
&=  \frac12 \tilde{C}^{ \dot{\alpha} \alpha} \bigg( -\hat{C}^{-1}_{\alpha \dot{\alpha} } (2\pi)^4\delta^{(4)}(0) + \frac{\delta^2 L_{\text{int.}}}{ \delta \overline{\chi}^{\dot{\alpha}}  \delta \chi^{\alpha}} - \frac{\delta L_{\text{prop.}} }{\delta \overline{\chi}^{\dot{\alpha}}} \frac{\delta L_{\text{prop.}}}{\delta \chi^{\alpha}} + \frac{\delta L_{\text{int.}}}{\delta \overline{\chi}^{\dot{\alpha}}} \frac{\delta L_{\text{int.}}}{\delta \chi^{\alpha}} \ + 
\\
& \qquad \qquad + \frac{\delta L_{\text{prop.}} }{\delta \overline{\chi}^{\dot{\alpha}}} \frac{\delta L_{\text{int.}}}{\delta \chi^{\alpha}} - \frac{\delta L_{\text{int.}}}{\delta \overline{\chi}^{\dot{\alpha}}} \frac{\delta L_{\text{prop.}} }{\delta \chi^{\alpha}} \bigg) \,,
\end{aligned}
\ee
where we see that the last lines of each of the above equations will cancel each other when added together; the first term in each equation can again be absorbed into the measure and the third terms, when summed up, will, by construction, cancel $\dot{L}^{(\chi)}_{\text{prop.}}$, which we have computed earlier. 
The invariance of the partition function requires 
\be
\dot L = - e^{-L} \left\{ \frac{\delta}{\delta \chi^\alpha} \left( \Psi_1^\alpha e^L \right) 
- \frac{\delta}{\delta \overline\chi^{\dot \alpha}} \left( \Psi_2^{\dot \alpha} e^L \right) \right\} \ + \ \text{bosonic contributions} \,, 
\ee
and therefore the final contribution to the flow of $L_{\text{int.}}$ is 
\be
\label{FERMFERM}
\begin{aligned}
\dot{L}_{\text{int.}} = - \int \frac{d^4 \hat{k}}{(2\pi)^4} \tilde{C}^{\dot{\alpha} \alpha }(\hat{k}) \bigg( \frac{\delta^2 L_{\text{int.}}}{ \delta \overline{\chi}^{\dot{\alpha}}(-\hat{k})  \delta \chi^{\alpha}(\hat{k})} + \frac{\delta L_{\text{int.}}}{\delta \overline{\chi}^{\dot{\alpha}}(-\hat{k})} \frac{\delta L_{\text{int.}}}{\delta \chi^{\alpha}(\hat{k})} \bigg) \ +\\+ \ \text{bosonic contributions} \,.
\end{aligned}
\ee
The $(2 \pi)^4$ in the denominator in \eqref{FERMFERM} appears, instead, in the numerator in \eqref{FermFermB} because, as we have already explained, in \eqref{FERMFERM} we have variational derivatives whereas in \eqref{FermFermB} we have partial derivatives, and they differ by a factor of $(2 \pi)^4$ when it comes to momentum matching. 

We make a final observation. For each massless chiral multiplet, which consists of a complex scalar field, a Weyl fermion and a complex auxiliary field, it is straightforward to check that the terms that would otherwise have to be absorbed into the measure in the right hand side of the ERG equation actually sum up to zero, which is a non trivial relation required by supersymmetry.


\begin{thebibliography}{99}


\bibitem{Kachru:2003aw}
S.~Kachru, R.~Kallosh, A.~D.~Linde and S.~P.~Trivedi,
``De Sitter vacua in string theory'',
Phys. Rev. D \textbf{68} (2003), 046005
[arXiv:hep-th/0301240 [hep-th]].




\bibitem{Balasubramanian:2005zx}
V.~Balasubramanian, P.~Berglund, J.~P.~Conlon and F.~Quevedo,
``Systematics of moduli stabilisation in Calabi-Yau flux compactifications'',
JHEP \textbf{03} (2005), 007
[arXiv:hep-th/0502058 [hep-th]].


\bibitem{Conlon:2005ki}
J.~P.~Conlon, F.~Quevedo and K.~Suruliz,
``Large-volume flux compactifications: Moduli spectrum and D3/D7 soft supersymmetry breaking'',
JHEP \textbf{08} (2005), 007
[arXiv:hep-th/0505076 [hep-th]].



\bibitem{Kallosh:2018nrk}
R.~Kallosh and T.~Wrase,
``dS Supergravity from 10d'',
Fortsch. Phys. \textbf{67} (2019) no.1-2, 1800071
[arXiv:1808.09427 [hep-th]].


\bibitem{Bento:2021nbb}
B.~V.~Bento, D.~Chakraborty, S.~L.~Parameswaran and I.~Zavala,
``A new de Sitter solution with a weakly warped deformed conifold'',
JHEP \textbf{12} (2021), 124
[arXiv:2105.03370 [hep-th]].


\bibitem{Bena:2022cwb}
I.~Bena, E.~Dudas, M.~Gra\~na, G.~L.~Monaco and D.~Toulikas,
``Bare-Bones de Sitter'',
[arXiv:2202.02327 [hep-th]].


\bibitem{Danielsson:2018ztv}
U.~H.~Danielsson and T.~Van Riet,
``What if string theory has no de Sitter vacua?'',
Int. J. Mod. Phys. D \textbf{27} (2018) no.12, 1830007
[arXiv:1804.01120 [hep-th]].


\bibitem{Obied:2018sgi} 
G.~Obied, H.~Ooguri, L.~Spodyneiko and C.~Vafa, 
``De Sitter Space and the Swampland'', 
[arXiv:1806.08362 [hep-th]].


\bibitem{Andriot:2018wzk} 
D.~Andriot, 
``On the de Sitter swampland criterion'', 
Phys. Lett. B \textbf{785} (2018), 570-573 
[arXiv:1806.10999 [hep-th]].


\bibitem{Garg:2018reu} 
S.~K.~Garg and C.~Krishnan, 
``Bounds on Slow Roll and the de Sitter Swampland'', 
JHEP \textbf{11} (2019), 075 
[arXiv:1807.05193 [hep-th]].


\bibitem{Ooguri:2018wrx}
H.~Ooguri, E.~Palti, G.~Shiu and C.~Vafa,
``Distance and de Sitter Conjectures on the Swampland'',
Phys. Lett. B \textbf{788} (2019), 180-184 
[arXiv:1810.05506 [hep-th]].


\bibitem{Moritz:2017xto}
J.~Moritz, A.~Retolaza and A.~Westphal,
``Toward de Sitter space from ten dimensions'',
Phys. Rev. D \textbf{97} (2018) no.4, 046010
[arXiv:1707.08678 [hep-th]].


\bibitem{Sethi:2017phn}
S.~Sethi,
``Supersymmetry Breaking by Fluxes'',
JHEP \textbf{10} (2018), 022
[arXiv:1709.03554 [hep-th]].


\bibitem{Hamada:2018qef}
Y.~Hamada, A.~Hebecker, G.~Shiu and P.~Soler,
``On brane gaugino condensates in 10d'',
JHEP \textbf{04} (2019), 008
[arXiv:1812.06097 [hep-th]].


\bibitem{Hamada:2019ack}
Y.~Hamada, A.~Hebecker, G.~Shiu and P.~Soler,
``Understanding KKLT from a 10d perspective'',
JHEP \textbf{06} (2019), 019
[arXiv:1902.01410 [hep-th]].


\bibitem{Carta:2019rhx}
F.~Carta, J.~Moritz and A.~Westphal,
``Gaugino condensation and small uplifts in KKLT'',
JHEP \textbf{08} (2019), 141
[arXiv:1902.01412 [hep-th]].


\bibitem{Gautason:2019jwq}
F.~F.~Gautason, V.~Van Hemelryck, T.~Van Riet and G.~Venken,
``A 10d view on the KKLT AdS vacuum and uplifting'',
JHEP \textbf{06} (2020), 074
[arXiv:1902.01415 [hep-th]].


\bibitem{Gao:2020xqh}
X.~Gao, A.~Hebecker and D.~Junghans,
``Control issues of KKLT'',
Fortsch. Phys. \textbf{68} (2020), 2000089
[arXiv:2009.03914 [hep-th]].


\bibitem{Junghans:2022exo}
D.~Junghans,
``LVS de Sitter Vacua are probably in the Swampland'',
[arXiv:2201.03572 [hep-th]].


\bibitem{Lindstrom:1979kq}
U.~Lindstrom and M.~Rocek,
``CONSTRAINED LOCAL SUPERFIELDS'',
Phys. Rev. D \textbf{19} (1979), 2300-2303


\bibitem{Kapustnikov:1981de}
A.~A.~Kapustnikov,
``NONLINEAR REALIZATION OF EINSTEINIAN SUPERGRAVITY'',
Theor. Math. Phys. \textbf{47} (1981), 406-413


\bibitem{Samuel:1982uh}
S.~Samuel and J.~Wess,
``A Superfield Formulation of the Nonlinear Realization of Supersymmetry and Its Coupling to Supergravity'',
Nucl. Phys. B \textbf{221} (1983), 153-177


\bibitem{Bergshoeff:2015tra}
E.~A.~Bergshoeff, D.~Z.~Freedman, R.~Kallosh and A.~Van Proeyen,
``Pure de Sitter Supergravity'',
Phys. Rev. D \textbf{92} (2015) no.8, 085040
[erratum: Phys. Rev. D \textbf{93} (2016) no.6, 069901] 
[arXiv:1507.08264 [hep-th]].


\bibitem{Cribiori:2017ngp}
N.~Cribiori, G.~Dall'Agata and F.~Farakos,
``From Linear to Non-linear SUSY and Back Again'',
JHEP \textbf{08} (2017), 117
[arXiv:1704.07387 [hep-th]].


\bibitem{DallAgata:2016syy}
G.~Dall'Agata, E.~Dudas and F.~Farakos,
``On the origin of constrained superfields'',
JHEP \textbf{05} (2016), 041
[arXiv:1603.03416 [hep-th]].


\bibitem{Ferrara:2014kva}
S.~Ferrara, R.~Kallosh and A.~Linde,
``Cosmology with Nilpotent Superfields'',
JHEP \textbf{10} (2014), 143
[arXiv:1408.4096 [hep-th]].


\bibitem{Bergshoeff:2015jxa}
E.~A.~Bergshoeff, K.~Dasgupta, R.~Kallosh, A.~Van Proeyen and T.~Wrase,
``$ \overline{\mathrm{D}3} $ and dS'',
JHEP \textbf{05} (2015), 058
[arXiv:1502.07627 [hep-th]].


\bibitem{Bandos:2015xnf}
I.~Bandos, L.~Martucci, D.~Sorokin and M.~Tonin,
``Brane induced supersymmetry breaking and de Sitter supergravity'',
JHEP \textbf{02} (2016), 080
[arXiv:1511.03024 [hep-th]].


\bibitem{Dasgupta:2016prs}
K.~Dasgupta, M.~Emelin and E.~McDonough,
``Fermions on the antibrane: Higher order interactions and spontaneously broken supersymmetry'',
Phys. Rev. D \textbf{95} (2017) no.2, 026003
[arXiv:1601.03409 [hep-th]].


\bibitem{Cribiori:2019bfx}
N.~Cribiori, R.~Kallosh, C.~Roupec and T.~Wrase,
``Uplifting Anti-D6-brane'',
JHEP \textbf{12} (2019), 171
[arXiv:1909.08629 [hep-th]].


\bibitem{Cribiori:2020bgt}
N.~Cribiori, C.~Roupec, M.~Tournoy, A.~Van Proeyen and T.~Wrase,
``Non-supersymmetric branes'',
JHEP \textbf{07} (2020), 189
[arXiv:2004.13110 [hep-th]].


\bibitem{Volkov:1973ix}
D.~V.~Volkov and V.~P.~Akulov,
``Is the Neutrino a Goldstone Particle?'',
Phys. Lett. B \textbf{46} (1973), 109-110.


\bibitem{Deser:1977uq}
S.~Deser and B.~Zumino,
``Broken Supersymmetry and Supergravity'',
Phys. Rev. Lett. \textbf{38} (1977), 1433-1436


\bibitem{Nambu:1961tp}
Y.~Nambu and G.~Jona-Lasinio,
``Dynamical Model of Elementary Particles Based on an Analogy with Superconductivity. 1.'',
Phys. Rev. \textbf{122} (1961), 345-358


\bibitem{Nambu:1961fr}
Y.~Nambu and G.~Jona-Lasinio,
``DYNAMICAL MODEL OF ELEMENTARY PARTICLES BASED ON AN ANALOGY WITH SUPERCONDUCTIVITY. II'',
Phys. Rev. \textbf{124} (1961), 246-254


\bibitem{Bardeen:1989ds}
W.~A.~Bardeen, C.~T.~Hill and M.~Lindner,
``Minimal Dynamical Symmetry Breaking of the Standard Model,''
Phys. Rev. D \textbf{41} (1990), 1647


\bibitem{Jasinschi:1983wr}
R.~S.~Jasinschi and A.~W.~Smith,
``Dynamical Mass Generation for the Gravitino in Simple $N=1$ Supergravity'',
Phys. Lett. B \textbf{173} (1986), 297-302


\bibitem{Jasinschi:1984cx}
R.~S.~Jasinschi and A.~W.~Smith,
``EFFECTIVE POTENTIAL IN N=1, d = 4 SUPERGRAVITY COUPLED TO THE VOLKOV-AKULOV FIELD'',
Phys. Lett. B \textbf{174} (1986), 183-185


\bibitem{Odintsov:1988wz}
S.~D.~Odintsov,
``Effective Potential in $N=1$ Supergravity De Sitter Space'',
Phys. Lett. B \textbf{213} (1988), 7-10


\bibitem{Buchbinder:1989gi}
I.~L.~Buchbinder and S.~D.~Odintsov,
``Is Dynamical Supersymmetry Breaking in $N=1$ Supergravity Possible?'',
Class. Quant. Grav. \textbf{6} (1989), 1955-1959


\bibitem{Ellis:2013zsa}
J.~Ellis and N.~E.~Mavromatos,
``Inflation induced by gravitino condensation in supergravity'',
Phys. Rev. D \textbf{88} (2013) no.8, 085029
[arXiv:1308.1906 [hep-th]].


\bibitem{Alexandre:2013iva}
J.~Alexandre, N.~Houston and N.~E.~Mavromatos,
``Dynamical Supergravity Breaking via the Super-Higgs Effect Revisited'',
Phys. Rev. D \textbf{88} (2013), 125017
[arXiv:1310.4122 [hep-th]].


\bibitem{Ishikawa:2019pnb}
R.~Ishikawa and S.~V.~Ketov,
``Gravitino condensate in $N=1$ supergravity coupled to the $N=1$ supersymmetric Born-Infeld theory'',
PTEP \textbf{2020} (2020) no.1, 013B05
[arXiv:1904.08586 [hep-th]].


\bibitem{Alexandre:2014lla}
J.~Alexandre, N.~Houston and N.~E.~Mavromatos,
``Inflation via Gravitino Condensation in Dynamically Broken Supergravity'',
Int. J. Mod. Phys. D \textbf{24} (2015) no.04, 1541004
[arXiv:1409.3183 [gr-qc]]. 


\bibitem{Houston:2015ygm}
N.~Houston,
``Gravitino condensation, supersymmetry breaking and inflation'',
[arXiv:1512.08210 [hep-th]].


\bibitem{Jaeckel:2002rm}
J.~Jaeckel and C.~Wetterich,
``Flow equations without mean field ambiguity'',
Phys. Rev. D \textbf{68} (2003), 025020
[arXiv:hep-ph/0207094 [hep-ph]].


\bibitem{Rocek:1978nb}
M.~Rocek,
``Linearizing the Volkov-Akulov Model'',
Phys. Rev. Lett. \textbf{41} (1978), 451-453.


\bibitem{Casalbuoni:1988xh}
R.~Casalbuoni, S.~De Curtis, D.~Dominici, F.~Feruglio and R.~Gatto,
``Nonlinear Realization of Supersymmetry Algebra From Supersymmetric Constraint'',
Phys. Lett. B \textbf{220} (1989), 569-575.


\bibitem{Komargodski:2009rz}
Z.~Komargodski and N.~Seiberg,
``From Linear SUSY to Constrained Superfields'',
JHEP \textbf{09} (2009), 066,
[arXiv: 0907.2441 [hep-th]].


\bibitem{Polchinski:1983gv}
J.~Polchinski,
``Renormalization and Effective Lagrangians'',
Nucl. Phys. B \textbf{231} (1984), 269-295.


\bibitem{Ball:1993zy}
R.~D.~Ball and R.~S.~Thorne,
``Renormalizability of effective scalar field theory'',
Annals Phys. \textbf{236} (1994), 117-204
[arXiv: hep-th/9310042 [hep-th]].


\bibitem{Zumbach:1994vg}
G.~Zumbach,
``The Renormalization group in the local potential approximation and its applications to the O(n) model'',
Nucl. Phys. B \textbf{413} (1994), 754-770


\bibitem{Zumbach:1994kc}
G.~Zumbach,
``The Local potential approximation of the renormalization group and its applications'',
Phys. Lett. A \textbf{190} (1994), 225-230


\bibitem{Morris:1994ie}
T.~R.~Morris,
``Derivative expansion of the exact renormalization group'',
Phys. Lett. B \textbf{329} (1994), 241-248
[arXiv:hep-ph/9403340 [hep-ph]].


\bibitem{Harvey-Fros:1999qpe}
C.~S.~F.~Harvey-Fros,
``The Local potential approximation of the renormalization group'',
[arXiv:hep-th/0108018 [hep-th]].


\bibitem{Litim:2018pxe}
D.~F.~Litim and M.~J.~Trott,
``Asymptotic safety of scalar field theories'',
Phys. Rev. D \textbf{98} (2018) no.12, 125006
[arXiv: 1810.01678 [hep-th]].


\bibitem{Fei:2016sgs}
L.~Fei, S.~Giombi, I.~R.~Klebanov and G.~Tarnopolsky,
``Yukawa CFTs and Emergent Supersymmetry'',
PTEP \textbf{2016} (2016) no.12, 12C105
[arXiv:1607.05316 [hep-th]].


\bibitem{Gies:2017tod}
H.~Gies, T.~Hellwig, A.~Wipf and O.~Zanusso,
``A functional perspective on emergent supersymmetry'',
JHEP \textbf{12} (2017), 132
[arXiv:1705.08312 [hep-th]].


\bibitem{Granda:1997xk}
L.~N.~Granda and S.~D.~Odintsov,
``Exact renormalization group for O(4) gauged supergravity'',
Phys. Lett. B \textbf{409} (1997), 206-212
[arXiv:hep-th/9706062 [hep-th]].


\bibitem{Percacci:2013ii}
R.~Percacci, M.~J.~Perry, C.~N.~Pope and E.~Sezgin,
``Beta Functions of Topologically Massive Supergravity'',
JHEP \textbf{03} (2014), 083
[arXiv:1302.0868 [hep-th]].


\bibitem{Kuzenko:2011tj}
S.~M.~Kuzenko and S.~J.~Tyler,
``On the Goldstino actions and their symmetries'',
JHEP \textbf{05} (2011), 055
[arXiv: 1102.3043 [hep-th]].


\bibitem{Cribiori:2016hdz}
N.~Cribiori, G.~Dall'Agata and F.~Farakos,
``Interactions of N Goldstini in Superspace'',
Phys. Rev. D \textbf{94} (2016) no.6, 065019
[arXiv:1607.01277 [hep-th]].


\bibitem{Rosten:2008ih}
O.~J.~Rosten,
``On the Renormalization of Theories of a Scalar Chiral Superfield'',
JHEP \textbf{03} (2010), 004
[arXiv:0808.2150 [hep-th]].


\bibitem{Brignole:2000kg}
A.~Brignole,
``One loop Kahler potential in non renormalizable theories'',
Nucl. Phys. B \textbf{579} (2000), 101-116
[arXiv: hep-th/0001121 [hep-th]].


\bibitem{Peskin:1995ev}
M.~E.~Peskin and D.~V.~Schroeder,
``An Introduction to quantum field theory'', 
Addison-Wesley (1995).  


\bibitem{Wess:1992cp}
J.~Wess and J.~Bagger,
``Supersymmetry and supergravity'', 
Princeton, USA: Univ. Pr.(1992). 


\bibitem{Feldmann:2017ooy}
P.~Feldmann, A.~Wipf and L.~Zambelli,
``Critical Wess-Zumino models with four supercharges in the functional renormalization group approach'',
Phys. Rev. D \textbf{98} (2018) no.9, 096005
[arXiv:1712.03910 [hep-th]].


\bibitem{Bervillier:2014tla}
C.~Bervillier, ``Structure of Exact Renormalization Group Equations for field theory'',
[arXiv:1405.0791 [hep-th]].

\bibitem{Kachru:2003sx}
S.~Kachru, R.~Kallosh, A.~D.~Linde, J.~M.~Maldacena, L.~P.~McAllister and S.~P.~Trivedi,
JCAP \textbf{10} (2003), 013
doi:10.1088/1475-7516/2003/10/013
[arXiv:hep-th/0308055 [hep-th]].

\bibitem{Bena:2018fqc}
I.~Bena, E.~Dudas, M.~Gra\~na and S.~L\"ust,
``Uplifting Runaways'',
Fortsch. Phys. \textbf{67} (2019) no.1-2, 1800100
[arXiv:1809.06861 [hep-th]].


\bibitem{Dudas:2019pls}
E.~Dudas and S.~L\"ust,
``An update on moduli stabilization with antibrane uplift'',
JHEP \textbf{03} (2021), 107
[arXiv:1912.09948 [hep-th]].


\bibitem{Farakos:2020wfc}
F.~Farakos, A.~Kehagias and N.~Liatsos,
``de Sitter decay through goldstino evaporation'',
JHEP \textbf{02} (2021), 186
[arXiv:2009.03335 [hep-th]].


\bibitem{Kachru:2002gs}
S.~Kachru, J.~Pearson and H.~L.~Verlinde,
``Brane / flux annihilation and the string dual of a nonsupersymmetric field theory'',
JHEP \textbf{06} (2002), 021
[arXiv:hep-th/0112197 [hep-th]].




\end{thebibliography}
\end{document}